\documentclass[prd,aps,floatfix,preprint,%
               superscriptaddress,showpacs,titlepage,%
               nofootinbib,titlepage,byrevtex]{revtex4}
\usepackage{graphicx}
\begin{document}

\newcommand{\Npoly}{N_{\mathit{poly}}}
\newcommand{\hatDA}{\hat{D}^{A}}
\newcommand{\hatDS}{\hat{D}^{S}}
\newcommand{\NMD}{N_{\mathrm{MD}}}
\newcommand{\NMult}{N_{\mathrm{Mult}}}

\date{\today}

%%%%%%%%% Title
\title{Polynomial hybrid Monte Carlo algorithm for lattice QCD with an 
       odd number of flavors}

%%%%%%%%% Address List
\newcommand{\Tsukuba}%
{Institute of Physics, University of Tsukuba, Tsukuba, Ibaraki 305-8571, Japan}%1

\newcommand{\RCCP}%
{Center for Computational Physics, University of Tsukuba, Tsukuba, Ibaraki 305-8577, Japan}%2

\newcommand{\ICRR}%
{Institute for Cosmic Ray Research, University of Tokyo, Kashiwa, Chiba 277-8582, Japan}%3

\newcommand{\Hiroshima}%
{Department of Physics, Hiroshima University, Higashi-Hiroshima, Hiroshima 739-8526, Japan}%33

\newcommand{\KEK}%
{High Energy Accelerator Research Organization(KEK), Tsukuba, Ibaraki 305-0801, Japan}%4

\newcommand{\YITP}%
{Yukawa Institute for Theoretical Physics, Kyoto University, Kyoto 606-8502, Japan}%5

%%%%%%%%% Author List
\author{S.~Aoki}
\affiliation{\Tsukuba}

\author{R.~Burkhalter}
\affiliation{\RCCP}

\author{M.~Fukugita}
\affiliation{\ICRR}

\author{S.~Hashimoto}
\affiliation{\KEK}

\author{K-I.~Ishikawa}
\affiliation{\RCCP}

\author{N.~Ishizuka}
\affiliation{\Tsukuba}
\affiliation{\RCCP}

\author{Y.~Iwasaki}
\affiliation{\Tsukuba}
\affiliation{\RCCP}

\author{K.~Kanaya}
\affiliation{\Tsukuba}
\affiliation{\RCCP}

\author{T.~Kaneko}
\affiliation{\KEK}

\author{Y.~Kuramashi}
\affiliation{\KEK}

\author{M.~Okawa}
\altaffiliation[Present address:]{
Department of Physics, Hiroshima University,
Higashi-Hiroshima, Hiroshima 739-8526, Japan}
%\affiliation{\Hiroshima}
\affiliation{\KEK}

\author{T.~Onogi}
\affiliation{\YITP}

\author{S.~Tominaga}
\affiliation{\RCCP}

\author{N.~Tsutsui}
\affiliation{\KEK}

\author{A.~Ukawa}
\affiliation{\Tsukuba}
\affiliation{\RCCP}

\author{N.~Yamada}
\affiliation{\KEK}

\author{T.~Yoshi\'{e}}
\affiliation{\Tsukuba}
\affiliation{\RCCP}

\collaboration{JLQCD Collaboration}
\noaffiliation

\preprint{KEK-CP-120}
\preprint{UTCPP-P-121}
\preprint{UTHEP-455}

%%%%%%%%% Abstract
\begin{abstract}
We present a polynomial hybrid Monte Carlo (PHMC) algorithm for
lattice QCD with odd numbers of flavors of $O(a)$-improved Wilson
quark action.
The algorithm makes use of the non-Hermitian Chebyshev polynomial
to approximate the inverse square root of the fermion matrix required
for an odd number of flavors.  The systematic error from the polynomial
approximation is removed by a noisy Metropolis test for which
a new method is developed.
Investigating the property of our PHMC algorithm in the $N_f=2$ QCD case,
we find that it is as efficient as the conventional HMC algorithm for
a moderately large lattice size ($16^{3}\times 48$) with
intermediate quark masses ($m_{\mathit{PS}}/m_{\mathit{V}}\sim 0.7$-$0.8$).
We test our odd-flavor algorithm through extensive simulations of
two-flavor QCD treated as an $N_f=1+1$ system, and comparing the results
with those of the established algorithms for $N_f=2$ QCD.
These tests establish that our PHMC algorithm works on a moderately
large lattice size with intermediate quark masses
($16^{3}\times 48, m_{\mathit{PS}}/m_{\mathit{V}}\sim 0.7$-$0.8$).
Finally we experiment with the $(2+1)$-flavor QCD simulation on small
lattices ($4^3\times 8$ and $8^3\times 16$), and confirm the agreement of
our results with those obtained with the $R$ algorithm and extrapolated to
a zero molecular dynamics step size.
\end{abstract}

\maketitle

%%%%%%%%%%%%%%%%%%%%%%%%%%%%%%%%%%%%%%%%%%%%%%%%%%%%%%%%%%%%%%%%%%%%%%%%%%%%%
\section{Introduction}

An essential step toward realistic lattice simulations of quantum
chromodynamics (QCD) is to develop efficient algorithms to incorporate 
the dynamical sea quark effects of up, down, and strange quarks. 
Most of the recent dynamical QCD simulations have been, however, 
limited to two-flavor QCD
where up and down quarks are treated dynamically while the loop effect
of the strange quark is still neglected.  This is mainly due to the lack of
efficient algorithms to treat an odd number of dynamical quark flavors.
The $R$ algorithm~\cite{Gottlieb_etal_Hybrid_R} is a possible candidate
for this purpose, but its serious drawback is the systematic error of
$O(dt^2)$ stemming from a finite step size $dt$ in the molecular dynamics
evolution. 
To control this systematic error, one has to keep $dt$ small enough and
to monitor the size of the error by performing simulations at 
various values of $dt$, which requires much computational effort.
Therefore, an exact algorithm such as the hybrid Monte Carlo (HMC) 
algorithm~\cite{Duane_et_al_HMC}, which is widely used for simulations with 
an even number of flavors, is clearly desirable.

Recently, Takaishi and de~Forcrand proposed an algorithm for an odd
number of dynamical flavors~\cite{Takaishi_Forcrand_Nf3}.
They use the polynomial hybrid Monte Carlo (PHMC)
algorithm~\cite{Forcrand_Takaishi_lat96,PHMC_Frezzotti_Jansen,PHMC_I_and_II_Frezzotti_Jansen}
with a non-Hermitian Chebyshev polynomial, with which one can
approximate the inverse square root of the fermion matrix needed for
the simulation of an odd number of
flavors~\cite{Borici_Forcrand,Nf_one_Alexandrou}. 
They introduced a method to calculate the correction factor required
to compensate for the truncation error due to the finite order of
the polynomial, and hence the algorithm is exact. 
The algorithm was tested on a small lattice for 1, $1+1$, and $2+1$
 flavors of Wilson fermions.

Clearly, the next step toward realistic simulations of QCD is to
investigate the practical feasibility of their algorithm 
for two light (up and down) quarks and one relatively heavy
(strange) quark on large physical volumes.
In this case, up and down quarks are treated with the usual pseudofermion 
method, while the strange quark is incorporated with the polynomial
approximation.
It is known that the multiboson algorithms, which also rely on
the polynomial approximation for the inverse of fermion matrix, 
fail for light quarks~\cite{Forcrand_MB_review}.
Therefore, we need to examine whether the algorithm with the polynomial
approximation works for intermediate quark masses (around the strange quark).
An implementation of the algorithm for the $O(a)$-improved Wilson (clover) quark
action~\cite{Sheikholeslami_Wohlert} is also important 
to carry out simulations with reduced systematic errors due to finite 
lattice spacing.

In this work we present a modified algorithm for $(2+1)$-flavor QCD with
the $O(a)$-improved Wilson quark action. 
Our algorithm is a variant of PHMC with the non-Hermitian Chebyshev 
polynomial as
that of Takaishi and de~Forcrand~\cite{Takaishi_Forcrand_Nf3}, while
the treatment of the correction factor is different.
We test our algorithm for two different systems.
One is two-flavor QCD treated as a system with $1+1$ flavors, 
and the simulation results are compared with those of the 
conventional HMC for two flavors.
The other is $(2+1)$-flavor QCD, where our algorithm is compared with the
$R$ algorithm~\cite{Gottlieb_etal_Hybrid_R} after extrapolating to zero
step size $dt\to 0$.

We also perform two systematic numerical tests of the HMC and PHMC 
algorithms in two-flavor QCD in order to provide a basis to find the 
best method and parameter choices for an extension to realistic
simulation with $2+1$ flavors.

As a first step, we test the even-odd preconditioning for the
$O(a)$-improved Wilson fermion action, which was first proposed by
Luo~\cite{Luo} and Jansen and Liu~\cite{Jansen_Liu}. 
They introduced symmetrical and asymmetrical preconditioning, and 
mainly considered the asymmetric version. 
In our practical tests we found significant improvement for both
versions over the simulation without preconditioning. 
The improvement is more pronounced for the symmetric case and the
computer time can be reduced almost by a factor two from that without
preconditioning.

Second, we investigate the efficiency of the PHMC algorithm depending
on the quark mass and on the degree of the polynomial. 
We found that the PHMC is as effective as the conventional HMC 
algorithm for two different quark masses corresponding to 
$m_{\mathit{PS}}/m_{\mathit{V}}=0.8$ and $0.7$ on a
reasonably large lattice.
This observation is encouraging, as it suggests that the polynomial
approximation is useful for future simulations of $(2+1)$-flavor QCD.

The rest of the paper is organized as follows.
In Sec.~\ref{sec:Outline_of_the_algorithm} we outline the algorithms 
we consider in this paper.
The polynomial hybrid Monte Carlo (PHMC) algorithm and its generalization 
to an odd number of flavors is described.
In Sec.~\ref{sec:Even_Odd_HMC} we test the efficiency of the
even-odd preconditioning for the $O(a)$-improved Wilson fermion action using
the usual HMC algorithm with two-flavor of quarks.
We then investigate the efficiency of the PHMC algorithm for
two-flavor QCD in Sec.~\ref{sec:PHMC}.
Section~\ref{sec:OddFlavor} describes details of our algorithm for 
an odd number of flavors, and presents some numerical tests
with which the consistency and the applicability is investigated.
Our conclusion is given in Sec.~\ref{sec:Conclusion}.
Our algorithm and simulation code have already been used for
a study of the phase structure of three-flavor QCD with the
Wilson-type fermion actions~\cite{JLQCD_1st_order}.

%%%%%%%%%%%%%%%%%%%%%%%%%%%%%%%%%%%%%%%%%%%%%%%%%%%%%%%%%%%%%%%%%%%%%%%%%%%%%%%%%
\section{Outline of the algorithm} 
\label{sec:Outline_of_the_algorithm}

We first present the outline of our algorithm for
$(N_{f_1}+N_{f_2})$-flavor QCD, where $N_{f_1}$ is an even number while
$N_{f_2}$ is odd.
The details of the algorithm will be explained separately in later
sections. 

In this section we consider the Wilson gauge and fermion actions, but
the algorithm can be applied to more complicated lattice actions
arising in the Symanzik improvement program~\cite{Symanzik}.
In particular, the algorithm is suitable for the $O(a)$-improved
Wilson action~\cite{Sheikholeslami_Wohlert} which has a clover-leaf-type 
operator to remove the discretization error of $O(a)$.

\subsection{Pseudofermion representation for even number of flavors}

Let $D_1$ and $D_2$ be the Dirac operators for two different fermion
masses corresponding to $N_{f_1}$ and $N_{f_2}$ flavors, respectively.
The partition function of this fermion system is given by 
\begin{equation}
 {\cal Z} = \int\!\! {\cal D}U\,
  (\det[D_1])^{N_{f_1}}
  (\det[D_2])^{N_{f_2}}
  e^{-S_{g}[U]},
  \label{eq:partition_function}
\end{equation}
where $S_g[U]$ represents the gauge action.

Since $N_{f_1}$ is an even number, the fermion determinant
$(\det[D_1])^{N_{f_1}}$ can be expressed in terms of the usual
pseudofermion integral 
\begin{equation}
 (\det[D_1])^{N_{f_1}}
  =
  \int\!\! {\cal D}\phi_1^{\dag} {\cal D}\phi_1
  \exp\left[
       - \left|D_1^{-N_{f_1}/2} \phi_1 \right|^2
     \right],
  \label{eq:det[D_1]}
\end{equation}
where we have used the relation $D_1^{\dag}=\gamma_5 D_1 \gamma_5$.
We use a short-hand notation for the norm of a vector $X$ as
$|X|^2 \equiv \sum_{n,\alpha,a}|X_{\alpha}^{a}(n)|^{2}$
with $n$ the site index, $\alpha$ the spinor index, and $a$ the color
index. 

In the usual HMC algorithm one uses some iterative solver to calculate
the inverse of the fermion matrix $D_1$.
In the PHMC
algorithm~\cite{PHMC_Frezzotti_Jansen,PHMC_I_and_II_Frezzotti_Jansen},
on the other hand, one introduces a polynomial $P_{\Npoly}[z]$ of
order $\Npoly$ that converges $1/z$ as $\Npoly \to \infty$. 
The non-Hermitian Chebyshev polynomial
\begin{equation}
 P_{\Npoly}[z] = \sum_{i=0}^{\Npoly} c_i (1-z)^i, 
  \label{eq:Chebyshev}
\end{equation}
with $c_i=(-1)^i$ is an example of such a polynomial, 
when $|1-z| < 1$.
Supposing that all eigenvalues of $D_1$ fall inside the complex domain
$|1-z| < 1$, we have
\begin{eqnarray}
 (\det[D_1])^{N_{f_1}}
  & = &
  \left[
   \frac{ \det[D_1 P_{\Npoly}[D_1]] }{
          \det[    P_{\Npoly}[D_1]] }
 \right]^{N_{f_1}}
  \nonumber \\
 & = &
  \left[
   \det[D_1 P_{\Npoly}[D_1]]
 \right]^{N_{f_1}}
  \int\!\!{\cal D}\phi_1^{\dag}{\cal D}\phi_1
  \exp\left[
       - \left|\left(P_{\Npoly}[D_1]\right)^{N_{f_1}/2}\phi_1
        \right|^2
     \right].
\label{eq:det[D_1]_polynomial}
\end{eqnarray}
We notice that the inversion of the fermion matrix
$D_{1}^{-N_{f_1}/2}$ is replaced by a calculation of the 
polynomial $\left(P_{\Npoly}[D_1]\right)^{N_{f_1}/2}$.

Following the original proposal of the multiboson algorithm by
L\"uscher~\cite{Luscher_MB}, 
Frezzotti and Jansen
\cite{PHMC_Frezzotti_Jansen,PHMC_I_and_II_Frezzotti_Jansen}
considered a Hermitian operator $Q = c_M\gamma_5 D_1$ with $c_M$ a
normalization factor and used a polynomial approximation of 
$\det[D_1]^2 = \det[Q]^2$ rather than the non-Hermitian $\det[D_1]$,
using the $\gamma_5$ Hermiticity property
$D_1^\dagger = \gamma_5 D_1 \gamma_5$ of the Wilson-type lattice
fermions.
In this work, however, we consider the non-Hermitian relation
Eq.~(\ref{eq:det[D_1]_polynomial}), as it is suitable for the extension 
to an odd number of flavors.

Since the polynomial approximation introduces a truncation error, one
has to evaluate the correction factor
$\left[\det[D_1 P_{\Npoly}[D_1]]\right]^{N_{f_1}}$
in order to make the algorithm exact.
As the correction factor is close to unity when the polynomial is a good 
approximation of the inverse, a stochastic technique can be used
to incorporate the correction factor.
The reweighting method~\cite{Luscher_MB} and 
the global Metropolis test~\cite{Borici_Forcrand,Borrelli_Forcrand_Galli} 
have been proposed and used in the multiboson algorithm.
For the PHMC algorithm, the reweighting method is applied in 
Refs.~\cite{PHMC_Frezzotti_Jansen,PHMC_I_and_II_Frezzotti_Jansen},
and the global Metropolis test in Ref.~\cite{Takaishi_Forcrand_Nf3}.
We use the global Metropolis test developed for a multiboson 
algorithm~\cite{Borici_Forcrand,Borrelli_Forcrand_Galli}. 
The details of the global Metropolis test in the case of $N_{f_1}=2$
will be given in Sec.~\ref{sub:NoisyMetropolis}.

\subsection{Pseudofermion representation for an odd number of flavors}
\label{sec:outline_odd_flavor}
For an odd number of flavors $N_{f_2}$, we use the method developed by
Alexandrou \textit{et al.}~\cite{Nf_one_Alexandrou} to take a ``square
root'' of the polynomial as described below.

We consider a polynomial $P_{\Npoly}[z]$ with an even degree $\Npoly$
and rewrite it as a product of monomials
\begin{equation}
 P_{\Npoly}[z] = \sum_{i=0}^{\Npoly} c_i (z-1)^i
  = c_{\Npoly} \prod_{k=1}^{\Npoly} (z-z_k),  
  \label{eq:Chebyshev2n}
\end{equation}
which approaches $1/z$ as $\Npoly$ increases.
At this point the convergence radius is assumed to cover all
eigenvalues of the Wilson-Dirac operator,
which will be confirmed in Sec.~\ref{sec:OddFlavor} numerically. 
Since $z_k$ appears with its complex conjugate,
we may rewrite Eq.~(\ref{eq:Chebyshev2n}) as 
\begin{equation}
 P_{\Npoly}[z] = c_{\Npoly} \prod_{j=1}^{\Npoly/2} (z-z_{k'(j)}^*)(z-z_{k(j)}),
\end{equation}
where $k(j)$ and $k'(j)$ are the arbitrary reordering indices
defined to satisfy the relation $z_{k(j)}^*=z_{k'(j)}^*$
with $j=1\ldots\Npoly/2$.
Using the property $D_2^\dagger = \gamma_5 D_2 \gamma_5$ one can show
that $\det[D_2-z_{k'(j)}^*] = \det[D_2-z_{k(j)}]^\dagger$ and 
\begin{eqnarray}
 \det[P_{\Npoly}[D_2]] & = &
  c_{\Npoly} \prod_{j=1}^{\Npoly/2} 
  \det[D_2-z_{k(j)}]^\dagger \det[D_2-z_{k(j)}]
  \nonumber\\
 & = &
  \det\left[
       T_{\Npoly}^\dagger[D_2] T_{\Npoly}[D_2]
     \right]
\label{eq:SplitdetD}
\end{eqnarray}
where 
$T_{\Npoly}[z]\equiv\sqrt{c_{\Npoly}}\prod_{j=1}^{\Npoly/2}(z-z_{k(j)})$.
Then we obtain a pseudofermion representation for an odd number of
flavors 
\begin{eqnarray}
 (\det[D_2])^{N_{f_2}} & = &
  \left[
   \frac{ \det[D_2 P_{\Npoly}[D_2]] }{\det[P_{\Npoly}[D_2]]}
 \right]^{N_{f_2}}
  \nonumber \\
 & = &
  \left[
   \det[D_2 P_{\Npoly}[D_2]]\right]^{N_{f_2}}
  \int\!\!{\cal D}\phi_2^{\dag}{\cal D}\phi_2
  \exp\left[
       -\left|
         \left(T_{\Npoly}[D_2]\right)^{N_{f_2}} \phi_2
       \right|^2
     \right].
  \label{eq:det[D_2]_odd}
\end{eqnarray}

As in the case of an even number of flavors, the correction factor 
$\left[\det[D_2 P_{\Npoly}[D_2]]\right]^{N_{f_2}}$ 
has to be kept to construct an exact algorithm.
We describe the calculation of the correction factor for the $N_{f_2}=1$ 
case in Sec.~\ref{sub:NoisyMetropolisOddFlavor}. 

We note that in this construction, the positivity of $\det[D_2]$ is assumed.
Since the Wilson-type lattice fermions does not have chiral
symmetry, the Wilson-Dirac operator $D_2$ may develop a
real and negative eigenvalue, which could make $\det[D_2]$ negative. 
In actual simulations, we do not expect that this happens for
the following reason.
Under a continuous change of gauge configuration, as in the molecular 
dynamics evolution, the eigenvalues also change continuously.
To change the sign of a real eigenvalue it has to cross zero, for
which the determinant $\det[D_2]$ vanishes which is suppressed. 
In addition, since the single flavor part is to be identified with 
strange quark in realistic applications, we expect that the intermediate 
mass of strange quark behaves as an infrared cutoff obstructing the 
appearance of negative eigenvalues.

In our implementation, we use the fact that the
correction factor $\left[\det[D_2 P_{\Npoly}[D_2]]\right]^{N_{f_2}}$ 
is close to unity.  If this does not hold, the calculation will fail
to converge.
We should, therefore, be aware of the appearance of a negative
determinant.
Our algorithm fails if this happens, but a negative
determinant should be considered as a problem of the formulation of
the lattice fermion rather than the problem of the algorithm, since it
is related to the lack of chiral symmetry.

\subsection{Hybrid Monte Carlo algorithm}
\label{subsec:HMC}
Once we write an effective action for the fermion determinant using 
pseudofermions as in
Eqs.~(\ref{eq:det[D_1]}), (\ref{eq:det[D_1]_polynomial}), and
(\ref{eq:det[D_2]_odd}), it is 
straightforward to apply the hybrid Monte Carlo
algorithm~\cite{Duane_et_al_HMC} to obtain an ensemble of gauge
configurations including the effect of the approximated fermion
determinant. 

Introducing a fictitious momentum $P$ conjugate to the link variable
$U$~(we suppress the site, direction, and color indices), 
the partition function Eq.~(\ref{eq:partition_function}) is written as 
\begin{equation}
 {\cal Z}=\int\!\!{\cal D}U{\cal D}P
  {\cal D}\phi_1^{\dag}{\cal D}\phi_1
  {\cal D}\phi_2^{\dag}{\cal D}\phi_2\det[W]e^{-H}.
  \label{eq:partition_function_H}
\end{equation}
If we use the usual form Eq.~(\ref{eq:det[D_1]}) for an even number of
flavors, and the polynomial representation Eq.~(\ref{eq:det[D_2]_odd}) for 
the rest of the fermions, the effective Hamiltonian $H$ and the
correction factor $\det[W]$ take the form
\begin{eqnarray}
 H & = & 
  \frac{1}{2} P^2 + S_g[U] 
  + |D_1^{-N_{f_1}/2}\phi_1|^2 
  + \left| 
     \left(T_{\Npoly}[D_2]\right)^{N_{f_2}} \phi_2
     \right|^2,
  \nonumber \\
 \det[W] & = &
  \left[\det[D_2 P_{\Npoly}[D_2]]\right]^{N_{f_2}}.
  \label{eq:hamiltonian_odd}
\end{eqnarray}
The HMC algorithm consists of the following four steps,
for a given gauge configuration $U$.
\begin{enumerate}
 \item[(1)] Generate momenta $P$ and pseudo-fermion fields $\phi_1$ and
            $\phi_2$ from a Gaussian distribution with unit variance and
            zero mean. 
 \item[(2)] Integrate link variables $U$ according to the discretized
            molecular dynamics evolution equation derived from the 
            equation of motion 
            \begin{eqnarray}
              \dot{U}_{\mu}(n) &=&   
              i  P_{\mu}(n) U_{\mu}(n), \nonumber \\
              \dot{P}_{\mu}(n) &=& 
            - i [U_{\mu}(n) F_{\mu}(n)]_{\mathrm{T.A.}},
            \end{eqnarray}
            where $\dot{X}$ is the derivative of a field $X$ with
            respect to the fictitious time $t$ and
            $[\cdots]_{\mathrm{T.A.}}$ means the traceless
            anti-Hermitian part of the matrix in the bracket. 
            The force $F_{\mu}(n)$ is defined through a variation of
            the effective Hamiltonian under an infinitesimal change
            $\delta U_{\mu}(n)$ of the gauge link variable
            \begin{equation}
              \label{eq:derivation_of_force}
              \delta H
              =
              \sum_{n,\mu}\mbox{Tr}\left[\{\delta U_{\mu}(n)F_{\mu}(n)\} 
                + \mbox{H.c.}\right].
            \label{eq:ForceDef}
            \end{equation}
            The length in the fictitious time $t$ is arbitrary, which we 
            set equal to unity throughout this paper.
 \item[(3)] Make a Metropolis test with respect to the energy difference
            $dH$ between the initial configuration $U(0)$
            and the trial configuration $U(t)$. 
            The acceptance probability is 
            $P_{acc}[(U(0),P(0)) \rightarrow (U(t),P(t))]
            = \mbox{min}[1,e^{-dH}]$.
            If the test is accepted go to the next step (4), or else 
            the new configuration is set to $(U(0),P(0))$ and go back to
            step (1). 
 \item[(4)] Make a Metropolis test with respect to the correction factor
            $\det[W]$. 
            If the test is accepted $(U(t),P(t))$ is taken as the new 
            configuration, or else the new configuration is $(U(0),P(0))$.
            Then return to step (1).
            The details to obtain the acceptance probability is described in
            Sec.~\ref{sub:NoisyMetropolisOddFlavor}.
\end{enumerate}

\section{Even-odd preconditioning for the $O(a)$-improved Wilson
 fermion action} 
\label{sec:Even_Odd_HMC}

Before going to the PHMC algorithm we discuss the
even-odd preconditioning of the fermion determinant. 
The even-odd preconditioning is a widely used technique to accelerate
the fermion matrix inversion~\cite{DeGrand_88}, but it can also be
used to reformulate the fermion determinant so that the pseudofermion
field lives only on odd sites~\cite{Gupta_et_al_89,DeGrand_Rossi}.
For the unimproved Wilson fermion action, no extra computational cost is
required by the reformulation, while the HMC simulation becomes
faster, since the phase space to be covered is reduced by a factor of two.
Luo~\cite{Luo} and Jansen and Liu~\cite{Jansen_Liu} introduced the even-odd
preconditioning for the $O(a)$-improved Wilson fermion which 
includes the clover-leaf-type operator. 
In this section we review their formulation and describe our extensive
numerical test to see how it improves the efficiency of the HMC
algorithm. 

\subsection{Description of the preconditioning}
The determinant of the $O(a)$-improved Wilson fermion operator $D$ 
is written as  
\begin{equation}
  \det[D]=\det\left(
    \begin{array}{cc}
      1+T_{ee} & M_{eo}\\
      M_{oe}   & 1+T_{oo}
    \end{array}
  \right),
  \label{eq:determinant}
\end{equation}
when the site index $n$ is numbered such that even sites come
earlier than any odd site.
Here, the site is even (odd), if $n_x+n_y+n_z+n_t$ is an even (odd)
number.
The hopping term $M$ ($M_{eo}$ or $M_{oe}$) represents the usual Wilson 
fermion matrix
\begin{equation}
 M_{n,n'} =-\kappa \sum_{\mu=1}^4
  \left\{
   (1-\gamma_\mu) U_\mu(n) \delta_{n+\hat{\mu},n'}
   +
   (1+\gamma_\mu) U_\mu^\dagger(n-\hat{\mu}) \delta_{n-\hat{\mu},n'}
  \right\}, 
\end{equation}
while $T$ ($T_{ee}$ or $T_{oo}$) describes the $O(a)$-improvement term
(or SW term)
\begin{equation}
 T_{n,n'} = 
  -\frac{1}{2}
  c_{\mathrm{sw}} \kappa \sigma_{\mu \nu} {\cal F}_{\mu \nu}(n)
  \delta_{n,n'},
\end{equation} 
with the clover-leaf-type field strength ${\cal F}_{\mu \nu}$ given by 
\begin{eqnarray}
 {\cal F}_{\mu \nu}(n) &=& 
  \frac{1}{8 i}
  \left[ \left\{
          U_{\mu}(n) U_{\nu}(n+\hat{\mu})
          U^{\dag}_{\mu}(n+\hat{\nu}) U^{\dag}_{\nu}(n)
 \right.\right.
  \nonumber\\
 & &
 + U_{\nu}(n) U^{\dag}_{\mu}(n+\hat{\nu}-\hat{\mu})
  U^{\dag}_{\nu}(n-\hat{\mu})U_{\mu}(n-\hat{\mu})
  \nonumber \\
 & &
  + U^{\dag}_{\mu}(n-\hat{\mu}) U^{\dag}_{\nu}(n-\hat{\mu}-\hat{\nu})
  U_{\mu}(n-\hat{\mu}-\hat{\nu}) U_{\nu}(n-\hat{\nu})
  \nonumber \\
 & &
  \left.\left.
         + U^{\dag}_{\nu}(n-\hat{\nu}) U_{\mu}(n-\hat{\nu})
         U_{\nu}(n-\hat{\nu}+\hat{\mu}) U^{\dag}_{\mu}(n) 
  \right\}
  - \mbox{H.c.}
 \right],
\end{eqnarray}
where $\mbox{H.c.}$ denotes the Hermitian conjugate of the preceding
bracket. 
The Dirac matrix $\gamma_\mu$ is defined such that it is Hermitian, and
$\sigma_{\mu \nu}=(i/2)[\gamma_{\mu},\gamma_{\nu}]$. 

Factoring out the even-even component ($1+T_{ee}$) from the determinant
Eq.~(\ref{eq:determinant}), we have 
\begin{equation}
 \det[D]=\det[1+T_{ee}]\det[\hatDA_{oo}], 
  \label{eq:AsymmetricDet}
\end{equation}
where
\begin{equation}
 \hatDA_{oo}=(1+T)_{oo}-M_{oe}(1+T)^{-1}_{ee} M_{eo}.
  \label{eq:AsymmetricDiracOp}
\end{equation}
It is also possible to factor out both the even-even and odd-odd
components as
\begin{equation}
 \det[D]= \det[1+T_{ee}]\det[1+T_{oo}]\det[\hatDS_{oo}],
  \label{eq:SymmetricDet}
\end{equation}
where
\begin{equation}
 \hatDS_{oo}=1-(1+T)^{-1}_{oo}M_{oe}(1+T)^{-1}_{ee} M_{eo}.
  \label{eq:SymmetricDiracOp}
\end{equation}
In the following, we refer to Eqs.~(\ref{eq:AsymmetricDet}) and 
(\ref{eq:SymmetricDet}) 
as asymmetric and symmetric preconditioning, respectively.
To our knowledge, previous simulations in the literature have
exclusively been made with the asymmetric even-odd preconditioning.

Using Eqs.~(\ref{eq:AsymmetricDet}) and (\ref{eq:AsymmetricDiracOp}), 
the asymmetrically preconditioned partition function for two flavor
QCD can be written as
\begin{eqnarray}
  {\cal Z}_{\mathrm{A\mbox{-}HMC}}
  &=&
  \int\!\!{\cal D}U{\cal D}P{\cal D}\phi_{o}^{\dag}{\cal D}\phi_{o}
  e^{-H_{\mathrm{A\mbox{-}HMC}}[P,U,\phi_o]},
  \nonumber \\
  H_{\mathrm{A\mbox{-}HMC}}[P,U,\phi_o]
  &=& 
  \frac{1}{2}P^{2} + S_{g}[U] 
  + S^{A}_q[U,\phi_o] + S^{A}_{\mathit{det}}[U],
  \nonumber \\
  S^{A}_q[U,\phi_o]
  &=&
  \left|\left(\hatDA_{oo}\right)^{-1}\psi_{o}\right|^2,
  \nonumber \\
  S^{A}_{\mathit{det}}[U]
  &=&
  -2 \log\det[1+T_{ee}].
  \label{eq:Asymmetric_EffAction}
\end{eqnarray}
The pseudofermion field $\phi_o$ lives on odd sites, whereas the
determinant $\det[1+T_{ee}]$ of the local SW term is calculated on
even sites. 

For the symmetrically  preconditioned partition function, from 
Eqs.~(\ref{eq:SymmetricDet}) and (\ref{eq:SymmetricDiracOp}) we have 
\begin{eqnarray}
  {\cal Z}_{\mathrm{S\mbox{-}HMC}}
  &=&
  \int\!\!{\cal D}U{\cal D}P{\cal D}\phi_{o}^{\dag}{\cal D}\phi_{o}
  e^{-H_{\mathrm{S\mbox{-}HMC}}[P,U,\phi_o]}, 
  \nonumber \\
  H_{\mathrm{S\mbox{-}HMC}}[P,U,\phi_o]&=& 
  \frac{1}{2}P^{2} + S_{g}[U] 
  + S^{S}_q[U,\phi_o] + S^{S}_{\mathit{det}}[U],
  \nonumber \\
  S^{S}_q[U,\phi_o]
  &=&
  \left|\left(\hatDS_{oo}\right)^{-1}\phi_{o}\right|^2,
  \nonumber \\
  S^{S}_{\mathit{det}}[U]
  &=&
  -2\left(\log\det[1+T_{ee}]+\log\det[1+T_{oo}]\right).
  \label{eq:Symmetric_EffAction}
\end{eqnarray}
In this case the determinant of the local SW term is calculated both on
even and odd sites. 

The calculation of the force defined in Eq.~(\ref{eq:derivation_of_force})
can be divided into several parts corresponding to the contribution
from the pure gauge action, the pseudofermion part, and the
determinant of the local SW term.
We write down the contribution from the quark part in
the Appendix for both preconditioning methods. 
%~\ref{sec:Appendix_A}

\subsection{Efficiency of the even-odd preconditioning}
The even-odd reformulation of the fermion determinant reduces the 
phase space to be covered by the HMC simulation. 
Another important effect of the preconditioning is that it lifts the
lowest eigenvalue of the fermion matrix and thus the
condition number is reduced.
The strength of the force coming from the pseudofermionic part
$S_q[U,\phi]$ of the effective Hamiltonian becomes
smaller~\cite{Forcrand_Takaishi_lat96},
since it is proportional
to the inverse of the lowest eigenvalue of the Dirac matrix.
Therefore, the error $dH$ accumulating in the molecular dynamics
evolution is also expected to become smaller, resulting in a better
acceptance rate in the HMC algorithm. 
To what extent the condition number is reduced depends on the
particulars of preconditioning.  We expect the symmetric one to work
better, since in the hopping parameter expansion $\hat{D}_{oo}^S$
behaves as $1-O(\kappa^2)$ while $\hat{D}_{oo}^A$ contains a
term proportional to $\kappa$ coming from $T_{oo}$.

In the following we describe a systematic test of the effect of the
preconditioning of both types.
The test is performed on three lattices:
(i) a small lattice of size $8^3 \times 16$ with a heavy quark mass,
which we call the ``small heavy'' lattice,    
(ii) a large lattice of size $16^3 \times 48$ with a heavy quark mass
called ``large heavy,'' and 
(iii) a large lattice of size $16^3\times 48$ with a light quark mass
called ``large light.'' 
Here, heavy and light quarks roughly correspond 
to $m_{\mathit{PS}}/m_{\mathit{V}}$=0.8
and 0.7, respectively.
The lattices (ii) and (iii) are reasonably large to study
the light hadron spectrum.  They are actually used in our production run
\cite{Kaneko_Lattice2000}. 
Details of the lattice parameters are listed in 
Table~\ref{tab:Lattice_Parameters}.

\subsection{Extensive test on a small lattice}
\label{sec:Test_on_SL}

On the small heavy lattice, we investigate the molecular dynamics 
(MD) step size 
$dt$ dependence of the acceptance rate $P_{\mathit{acc}}$ for
each algorithm:  
``HMC'' denotes the HMC algorithm without the preconditioning, 
``A-HMC'' and ``S-HMC'' are used for the asymmetrically or symmetrically
preconditioned HMC algorithm. 

We employ the BiCGStab algorithm \cite{BiCGStab} to calculate the
inverse of the Dirac matrix $D$ (or $\hatDA_{oo}$, $\hatDS_{oo}$). 
The symmetrical even-odd preconditioning is applied in the solver to 
accelerate the convergence of inversion. 
The stopping condition is defined so that the solver iterates 
until the residual defined by $r \equiv \sqrt{|D x - b|^2/|b|^{2}}$ 
becomes smaller than a certain value, where $b$ is a source
vector and $x$ is the solution vector. 
On the small heavy lattice, we use a rather strict stopping
condition to avoid systematic errors coming from the matrix
inversion. 
All numerical calculations are made with the double precision (64 bit)
arithmetic.
In Table~\ref{tab:Small_Heavy_Parameters}, we show the number of the
molecular dynamics (MD) steps $\NMD$ ($dt=1/\NMD$) and the stopping
condition for the BiCGStab solver in force and Hamiltonian
calculations. 
 
For the MD evolution of the kinematical variables $U$ and $P$, the
simplest integration scheme to satisfy the reversibility and
measure preservation is the leapfrog algorithm.
In this work we first consider two options of the leapfrog algorithm, 
i.e., $UPU$ and $PUP$ integrators.
In the $UPU$ integrator, the link variable $U$ is updated at the first
half step and then the integration of $P$ with a unit step size  
$dt$ follows.
Thus the link variable $U$ is assigned at $(n+1/2)\cdot dt$ 
with an integer $n$, while $P$ is assigned at $n \cdot dt$.
The integration is performed in the reverse order in the $PUP$
integrator. 

The acceptance rate in the HMC algorithm is governed by a change of
the effective Hamiltonian during the MD evolution 
$\langle dH\rangle$ as 
$P_{\mathit{acc}} = {\mathrm{erfc}}(\langle dH\rangle^{1/2}/2)$.
With the leapfrog integrator the change of effective Hamiltonian
behaves as $\langle dH\rangle \sim dt^4$ for small
$dt$~\cite{Creutz_88,Gupta_Kilcup_Sharpe_88,Gupta_Irback_Karsch_Petersson_90}. 

In Fig.~\ref{fig:dHvsdt_UPUvsPUP} we show the MD step size $dt$
dependence of $\langle dH\rangle$ for both $UPU$ and $PUP$ integrators.
The dotted lines represent a fit with a form
$\langle dH\rangle=\pi (a \cdot dt)^4$.
The Metropolis acceptance rate is plotted in
Fig.~\ref{fig:Paccvsdt_UPUvsPUP} as a function of $dt$. 
The expected behavior 
${\mathrm{erfc}}[\sqrt{\pi}(a \cdot dt)^{2}/2]$ is also shown by dotted
curves. 
We observe that the data is described by the expected functional
form. 
We also find that the $UPU$ integrator gives better acceptance
at a fixed $dt$ than the $PUP$ integrator, which has been known for
a long time for the staggered fermion action~\cite{Gupta_Kilcup_Sharpe_88}.
The computational cost with the $UPU$ integrator is lower by a
factor $\NMD/(\NMD+1)$ than the $PUP$ integrator since 
the computer time in dynamical QCD simulations is dominated 
by the force calculation that involves the fermion matrix inversion.
Therefore the advantage of the $UPU$ integrator is very clear.
We then use the $UPU$ integrator in the rest of this work.

Let us now discuss the effect of preconditioning.
Figures~\ref{fig:dHvsdt} and \ref{fig:Paccvsdt} show the MD step size
dependence of $\langle dH\rangle$ and  $P_{\mathit{acc}}$ for the HMC,
A-HMC and S-HMC algorithms. 
The $dt$ dependence for each algorithm is described very well by the 
relation $\langle dH\rangle\propto dt^4$ as shown in
Fig.~\ref{fig:dHvsdt}, and the value of $\langle dH\rangle$ for
A-HMC (S-HMC) at a fixed $dt$ is about a factor 5 (13) smaller than the
unpreconditioned HMC.
As a result, the acceptance is greatly improved as shown in
Fig.~\ref{fig:Paccvsdt}.
For instance, at $dt$=0.02 $P_{\mathit{acc}}$ is 81\% (88\%) for A-HMC
(S-HMC) compared to 60\% for the unpreconditioned case.

The efficiency of the algorithm may be defined as
$P_{\mathit{acc}} \cdot dt$ following Ref.~\cite{Takaishi_Choice_Integ}.
In order to plot the efficiency $P_{\mathit{acc}} \cdot dt$
as a function of $P_{\mathit{acc}}$,
we make use of an approximation of $P_{\mathit{acc}}$:
\begin{equation}
  P_{\mathit{acc}}
  = \exp\left(-(a\cdot dt)^2 - \frac{1}{2}(a\cdot dt)^4 \right).
  \label{eq:PaccvsdtApprox}
\end{equation}
This approximation is valid for small $dt$ [up to $O(dt^6)$]
and $\langle dH\rangle=\pi(a\cdot dt)^4$.
The validity can be ascertained in Fig.~\ref{fig:Paccvsdt},
where the approximation Eq.~(\ref{eq:PaccvsdtApprox}) is plotted (dotted curve) 
as well as the exact one ${\mathrm{erfc}}[\sqrt{\pi}(a\cdot dt)^{2}/2]$ (dashed
curve). 
Solving Eq.~(\ref{eq:PaccvsdtApprox}) for $dt$,
we obtain the explicit functional form for the efficiency 
$P_{\mathit{acc}} \cdot dt$ as
\begin{equation}
  P_{\mathit{acc}}\cdot dt =
  \frac{P_{\mathit{acc}}}{a}
  \sqrt{\sqrt{1-2\log(P_{\mathit{acc}})}-1},
  \label{eq:PaccdtvsPaccApprox}
\end{equation}
where the only parameter is $a$ defined through 
$\langle dH\rangle = \pi(a\cdot dt)^4$.
In Fig.~\ref{fig:EffvsPacc} we plot the efficiency 
$P_{\mathit{acc}}\cdot dt$ as a function of $P_{\mathit{acc}}$,
and Eq.~(\ref{eq:PaccdtvsPaccApprox}) is plotted as a dotted line.
It is remarkable that the optimal efficiency is reached when
$P_{\mathit{acc}}\simeq$ 0.65 irrespective of details of the
algorithm as far as we use the simplest leapfrog integrator for the MD 
evolution~\cite{Takaishi_Choice_Integ}.~\footnote{
  In Ref.~\cite{Takaishi_Choice_Integ} the maximum efficiency is
  reached at $P_{\mathit{acc}}\simeq$ 0.61 rather than 0.65.
  This difference comes from the expansion Eq.~(\ref{eq:PaccvsdtApprox})
  of the $\mathrm{erfc}$ function: the author of Ref.~\cite{Takaishi_Choice_Integ}
  considered the lowest order only, while we include the second order.
}
The efficiency of the algorithm can be measured by the parameter $a$. 
We therefore conclude that the efficiency of the A-HMC is 
a factor 1.5 better than the unpreconditioned HMC on the ``small
heavy'' lattice, and that of S-HMC by a factor of 1.9 which is even 
better.  

\subsection{Reversibility}
\label{sec:Reversibility}

Before we extend the comparison of the preconditioning to the large
($16^3\times 48$) lattice, we describe our choice of the stopping
condition for the Wilson-Dirac operator inverter on the large lattice, 
since it is computationally not realistic to keep the very strict
conditions of Sec.~\ref{sec:Test_on_SL} for the large lattice size.
The stopping condition in the calculation of the force may be relaxed
as far as the reversibility condition is maintained, which is tested
in the following.
In this section we employ the ``S-HMC'' preconditioning
to investigate the reversibility.

As a measure of how far one may loosen the stopping condition, we use the
violation of the reversibility condition for the effective Hamiltonian
defined by
\begin{equation}
  |\Delta H| =|H(t_r-t_r)-H(0)|,
  \label{eq:Def_Er_Hamil}
\end{equation}
where $H(t_r-t_r)$ means the effective Hamiltonian calculated 
for the reversed configuration which is obtained from the initial
configuration at $t=0$ by integrating the equation of motion to $t=t_r$
and then integrating back to $t=0$. The length of trajectory is $t_r$=1.
For the S-HMC effective action, we measure $|\Delta H|$ for several
values of the stopping condition on 20 thermalized configurations
separated by 10 trajectories. 
Figures~\ref{fig:DHDUDPrev_LH_HMC} and \ref{fig:DHDUDPrev_LL_HMC}
show $\langle |\Delta H|/H\rangle$ measured on the ``large heavy'' and
``large light'' lattices, respectively.
While the violation stays around the limit of the double precision 
arithmetic for the heavy dynamical quark
(Fig.~\ref{fig:DHDUDPrev_LH_HMC}), it depends on the stopping
condition for the light dynamical quark
(Fig.~\ref{fig:DHDUDPrev_LL_HMC}). 

The behavior for the light quark mass can be understood as follows. 
If the initial vector in the BiCGStab solver is reversible
($x=b$ is adopted in this work), the only source of the reversibility
violation is the round-off errors in the numerical computation.
Therefore, the error accumulates as the BiCGStab solver iterates and
thus the violation increases as the stopping condition is tightened.
This can be seen in Fig.~\ref{fig:DHDUDPrev_LL_HMC} from 
$r=10^{-5}$ to $10^{-7}$.
As we further decrease the stopping condition, the BiCGStab solver
gives a solution vector with better accuracy, and the value of
$\langle |\Delta H|/H\rangle$ is governed by the accuracy of the
solution vector.
It decreases as we tighten the stopping condition from
$r=10^{-7}$ to $10^{-13}$.  

As criteria to choose the stopping condition, we demand that the
solver iterates to the region where the 
$\langle|\Delta H|/H\rangle$ is governed by the accuracy of 
the solution vector and that the variation of the Hamiltonian over 
the trajectory $dH$ is not distorted by the error of the
solution vector.
These criteria are satisfied for $r\leq 10^{-7}$, and we choose
$10^{-8}$ in the following simulations in this work.

For completeness, we also calculate the violation of 
reversibility in the link variables $U$ and the conjugate momenta $P$
\begin{eqnarray}
  |\Delta U| &=&
  \sqrt{
    \sum_{n,\mu,a,b}|(U_{\mu})_{a,b}(n)(t_r-t_r)-(U_{\mu})_{a,b}(n)(0)|^2},
  \nonumber \\
  |\Delta P| &=&
  \sqrt{
    \sum_{n,\mu,a,b}|(P_{\mu})_{a,b}(n)(t_r-t_r)-(P_{\mu})_{a,b}(n)(0)|^2}, 
  \label{eq:Def_RelErr_DUDP}
\end{eqnarray}
where the sum runs over all sites $n$, color indices $a$, $b$, and
vector index $\mu$.
The results for $|\Delta U|$ and $|\Delta P|$ normalized by
$\sqrt{9\times 4\times N_{\mathrm{vol}}}$ with $N_{\mathrm{vol}}$ 
the total number of lattice site
are also plotted in
Figs.~\ref{fig:DHDUDPrev_LH_HMC} and \ref{fig:DHDUDPrev_LL_HMC},
where we observe the same pattern of the stopping condition dependence
as that of $\langle|\Delta H|/H\rangle$.

Since the MD evolution is chaotic, the violation of reversibility due
to the rounding error may grow
exponentially~\cite{Edwards_Horvath_Kennedy,Liu_Jaster_Jansen}.
The UKQCD Collaboration studied the reversibility for the same lattice
action as ours (but with the asymmetric preconditioning) with similar
lattice parameters.
They confirmed the exponential instability when the stopping
condition is too loose~\cite{UKQCD_Reversibility_Joo_etal}. 
The stopping condition we adopt $r<10^{-8}$ is strict
enough and no such problem emerges in our case. 
We also note that in Ref.~\cite{UKQCD_Reversibility_Joo_etal} most of the
numerical calculation is made with the single precision (32 bits)
arithmetic, while we use double precision throughout this work.

For the stopping condition in the Hamiltonian calculation, we keep 
a strict condition $r<10^{-14}$, since there is a
large cancellation in the difference $dH=H(t_r)-H(0)$, and the
accuracy of $dH$ is essential for the Metropolis test to be correct.

\subsection{Efficiency on large lattices}

We list the simulation parameters used for A-HMC and S-HMC algorithms
in Tables~\ref{tab:Large_Heavy_Parameters_HMC} (``large heavy'') and
\ref{tab:Large_Light_Parameters_HMC} (``large light'').
HMC means without preconditioning.
We observe that the number of MD steps is much reduced
for the preconditioned HMC algorithm compared to the unpreconditioned
one at the almost same acceptance rate. 
More precisely,
using the relation $\langle dH\rangle = \pi(a dt)^{4}$, we can compare
the best efficiency of the algorithm which depends only on $a$
as in Eq.~(\ref{eq:PaccdtvsPaccApprox}).
The gain is $1.5$  from HMC to A-HMC ($1.9$ from HMC to S-HMC) 
on the large heavy lattice.
For the large light lattice it is $1.8$ from HMC to A-HMC ($2.2$ from HMC to S-HMC).
The effect of even-odd preconditioning becomes more significant for lighter
quark masses (note that for infinite quark mass there is no choice for even-odd
preconditioning and is no improvement).
From our tests we conclude that the symmetrically even-odd 
preconditioning is again the best choice within the simple 
even-odd preconditioning.
Further improvement of the HMC algorithm may be achieved by 
preconditioning the partition function with incomplete LU factorization
type preconditioning~\cite{DeGrand_Rossi,Forcrand_Takaishi_lat96,Peardon,Hasenbusch}.
Hereafter we employ the symmetrically even-odd preconditioned form for the
$O(a)$-improved Wilson-Dirac operator and the QCD partition function
to develop the PHMC algorithm.

%%%%%%%%%%%%%%%%%%%%%%%%%%%%%%%%%%%%%%%%%%%%%%%%%%%%%%%%%%%%%%%%%%%%%%%%%%%%%%
\section{PHMC algorithm for two-flavor QCD}
\label{sec:PHMC}

The use of the polynomial hybrid Monte Carlo (PHMC) algorithm is
essential for the construction of an exact algorithm for odd number of
flavors.
Before going to the PHMC algorithm with odd number of quarks,
we investigate and develop the PHMC algorithm with two-flavors of quarks.
The PHMC algorithm in two-flavor QCD
was first proposed by Frezzotti and Jansen 
\cite{PHMC_Frezzotti_Jansen,PHMC_I_and_II_Frezzotti_Jansen}.
Some numerical tests of its performance were made 
\cite{PHMC_I_and_II_Frezzotti_Jansen,ALPHA_benchmark}
on small lattices and used for the determination of
$c_{\mathrm{sw}}$~\cite{ALPHA_csw} or the running coupling
constant~\cite{ALPHA_coupling} with the Schr\"odinger functional method.
In this section we perform further tests to explore the most
effective choice of the polynomial and its degree with the PHMC algorithm
in two-flavor QCD.
The numerical simulation and its comparison to the HMC algorithm are
carried out with the same lattice parameters employed in 
Sec.~\ref{sec:Even_Odd_HMC}.

\subsection{Partition function}

The partition function of two-flavor QCD with the symmetrical
preconditioning in the PHMC algorithm is given by 
\begin{eqnarray}
  {\cal Z}_{\mathrm{PHMC}}
  &=&
  \int\!\!{\cal D}U {\cal D}P {\cal D}\phi_o^{\dag}{\cal D}\phi_o
  \left(\det[W_{oo}]\right)^2
  e^{-H_{\mathrm{PHMC}}[P,U,\psi_o]},
  \nonumber\\
  W_{oo}
  &=&
  \hatDS_{oo} P_{\Npoly}[\hatDS_{oo}], 
  \nonumber \\
  H_{\mathrm{PHMC}}[P,U,\phi_o]
  &=&
  \frac{1}{2}P^2 + S_{g}[U]  
  + S^{S}_{\mathit{poly}}[U,\phi_o] 
  + S^{S}_{\mathit{det}}[U],
  \nonumber \\
  S^{S}_{\mathit{poly}}[U,\phi_o]
  &=&
  \left|P_{\Npoly}[\hatDS_{oo}]\phi_{o}\right|^{2},
  \label{eq:partition_function_H_PHMC}
\end{eqnarray}
and $S^{S}_{\mathit{det}}[U]$ is the same as in
Eq.~(\ref{eq:Symmetric_EffAction}).
The difference from Eq.~(\ref{eq:Symmetric_EffAction}) is in the
pseudofermion action $S_{\mathit{poly}}^{S}[U,\phi]$  
and in the insertion of the correction factor
$\left(\det[W_{oo}]\right)^2$. 

The polynomial $P_{\Npoly}[z]=\sum_{i=0}^{\Npoly} c_{i}(z-1)^{i}$
approximates $1/z$ for a complex $z$ placed in the convergence
region as $\Npoly \to \infty$,
and the coefficients $c_{i}$ are chosen so as to make the correction
factor $\left(\det[W_{oo}]\right)^2$ as close to unity as possible 
for small $\Npoly$.
For this purpose, several polynomials have been investigated in the
literature.
They include Chebyshev~\cite{Borici_Forcrand},
least-squares~\cite{Montvay_LeastSqueare}, and
adopted (with or without the UV
filtering)~\cite{Forcrand_UV_adopted,Alexandrou_UV_filter} 
polynomials.  
We consider the Chebyshev and adopted polynomials in this work.

The Chebyshev polynomial with unit circle convergence domain 
in the complex plane is defined by $c_{i}=(-1)^{i}$.
This is the same as the Taylor expansion with respect to the hopping
matrix. 
We call the PHMC algorithm with the Chebyshev polynomial as C-PHMC.
In this case, the accuracy of the polynomial is characterized by
$|zP_{\Npoly}[z]-1|=(z-1)^{\Npoly+1}$.

The coefficients $c_{i}$ for the adapted polynomial are determined so 
as to minimize 
$\left|\hatDS_{oo} P_{\Npoly}[\hatDS_{oo}]\eta_{o} - \eta_{o}\right|^{2}$
with a Gaussian noise vector $\eta_{o}$ with unit variance on a
thermalized background gauge
configuration~\cite{Forcrand_UV_adopted}. 
The coefficients thus obtained do neither show a strong dependence on 
the background gauge configuration nor on the noise vector.
We call this choice as A-PHMC.
We note that the adapted polynomial with the UV filtering is simple
and proven to be more efficient for the unimproved Wilson fermion in the
multiboson algorithm~\cite{Forcrand_UV_adopted,Alexandrou_UV_filter}.
For the $O(a)$-improved Wilson fermion, on the other hand, the UV
filtering requires an additional term in the effective Hamiltonian,
and to our knowledge its efficiency has not been tested yet.
We leave it as a future subject to study the efficiency of the UV filtering
for the PHMC algorithm with the $O(a)$-improved Wilson fermion action.

\subsection{Force calculation}
\label{subsec:Force_calculation}

The molecular dynamics step in the PHMC algorithm requires a
calculation of the force 
$\delta S^{S}_{\mathit{poly}}[U,\phi_o]/\delta U_\mu(n)$.
Frezzotti and
Jansen~\cite{PHMC_Frezzotti_Jansen,PHMC_I_and_II_Frezzotti_Jansen}
proposed to use a product representation of the polynomial
\begin{equation}
  P_{\Npoly}[z]
  = \sum_{i=0}^{\Npoly} c_{i}(z-1)^{i}
  = c_{\Npoly}\prod_{k=1}^{\Npoly}(z-z_k),
  \label{eq:PolyProdDef}
\end{equation}
with $z_{k}$ the roots of $P_{\Npoly}[z]=0$.
The computational cost can be reduced by holding the intermediate
vectors obtained by multiplying monomials on the pseudofermion field.

In the product representation with a naive ordering of monomials,
however, there is a problem of numerical instability and accumulation
of round-off errors due to the fact that the partial product
$\prod_{k=1}^l(z-z_k)$ fluctuates by several orders of magnitude for
intermediate $l$~\cite{Bunk_et_al_Ordering}.
The problem becomes more severe for small $z$ and for large order of the
polynomial. 
Bunk \textit{ et al.} proposed some ordering schemes to 
minimize the problem~\cite{Bunk_et_al_Ordering}.
In this work, instead of the product representation, we consider the
Clenshaw's recursive relation~\cite{Numerical_Recipe}
\begin{equation}
  P_{\Npoly}[z]
  % = \sum_{i=0}^{\Npoly} c_{i}(z-1)^{i} \\
  = c_0
  \left[1+\frac{c_1}{c_0}(z-1)
    \left[1+\frac{c_2}{c_1}(z-1)
      \left[ \cdots
        \left[1+\frac{c_{\Npoly}}{c_{\Npoly-1}}(z-1)\right] 
    \cdots \right] \right] \right],
  \label{eq:PolySumDef}
\end{equation}
for which the numerical instability is suppressed, because we 
add from the term giving the smallest contribution to the
terms with larger contributions.
We assume that the polynomial is converging and the higher order terms
are smaller; otherwise the algorithm does not work efficiently.

Adopting this representation, we expect that the calculation of the 
pseudofermion action
$S^{S}_{\mathit{poly}}$ in Eq.~(\ref{eq:partition_function_H_PHMC})
before and after the MD evolution
become stable numerically.
The force from $\delta S^{S}_{\mathit{poly}}[U,\phi_o]$ is given by
\begin{equation}
  \delta S^{S}_{\mathit{poly}}= 
  \left\{
    \sum_{j=1}^{\Npoly}\left[
      \left.X^{P(j)}\right.^{\dag} \delta M Y^{P(j)}
      +\left.X^{P(j)}\right.^{\dag} \delta T Z^{P(j)} \right]
  \right\} + \mbox{H.c.},
  \label{eq:force_from_polynomial}
\end{equation}
where
\begin{eqnarray}
 X^{P(j)}&=&\left(
               \begin{array}{c}
                   -(1+T)_{ee}^{-1} M_{oe}^{\dag} \hat{X}^{P(j)}_{o}\\
                   \hat{X}^{P(j)}_{o}
               \end{array}
            \right),\\
 Y^{P(j)}&=&\left(
               \begin{array}{c}
                   -(1+T)_{ee}^{-1} M_{eo} \hat{Y}^{P(j)}_{o}\\
                   \hat{Y}^{P(j)}_{o}
               \end{array}
            \right), \\
 Z^{P(j)}&=&\left(
               \begin{array}{c}
                   -(1+T)_{ee}^{-1} M_{eo} \hat{Y}^{P(j)}_{o}\\
                   (1+T)_{oo}^{-1}M_{oe}(1+T)_{ee}^{-1}M_{eo} \hat{Y}^{P(j)}_{o}
               \end{array}
            \right), \\
\hat{X}^{P(j)}_{o}&=&
              (1+T)_{oo}^{-1}
              \left[ \left(\hatDS_{oo}-1\right)^{\dag} \right]^{j-1}\hat{Y}^{P(0)}_{o}\\
\hat{Y}^{P(j)}_{o}&=& 
               c_{j}\left[ 1 + \frac{c_{j+1}}{c_{j}}  \left(\hatDS_{oo}-1\right)
   \right. \nonumber \\
   &&\times
                    \left[ 1 + \frac{c_{j+2}}{c_{j+1}}\left(\hatDS_{oo}-1\right)
   \right. \nonumber \\
   &&\times
                    \left[ 1 + \frac{c_{j+3}}{c_{j+2}}\left(\hatDS_{oo}-1\right)
   \right. \nonumber  \\   
   && \vdots \nonumber \\
   &&\times 
   \left. \left. \left.
            \left[ 1 + \frac{c_{\Npoly}}{c_{\Npoly-1}}\left(\hatDS_{oo}-1\right)\right]
      \cdots \right]
   \right]\right] \phi_{o}.
\label{eq:Force_PHMC_even}
\end{eqnarray}
In our implementation of the simulation code, we first calculate
$\hat{Y}^{P(j)}_{o}$ from $j=\Npoly$ to $0$ and store them on memory.  
We then calculate $\hat{X}^{P(j)}_{o}$ and the force
from $j=1$ to $\Npoly$ using the stored $\hat{Y}^{P(j)}_{o}$.
We do not need to store  $\hat{X}^{P(j)}_{o}$.
The requirement for memory is therefore the same as in the product
representation used
in Refs.~\cite{PHMC_Frezzotti_Jansen,PHMC_I_and_II_Frezzotti_Jansen}.

A potential source of the round-off errors in the calculation of the
force is the sum over $j$ in Eq.~(\ref{eq:force_from_polynomial}),
because the sum runs from the highest order to the lowest order in $\kappa$,
the $j$th term being of order $\kappa^{2(j-1)}$.
The numerical problem in the calculation of the force may be checked
by looking at the violation of reversibility.
We expect that the reversibility violation is small compared to HMC
because the MD evolution involves no iterative processes and is completely 
deterministic.
Numerical stability of the summation representation 
of polynomial and the reversibility of the molecular dynamics
will be discussed in 
Sec.~\ref{sec:PHMC_on_large_lattices}.

The pseudofermion field $\phi_{o}$ is generated from the Gaussian
noise vector $\eta_{o}$ at the beginning of the molecular dynamics
step through 
\begin{equation}
  \phi_{o} = P_{\Npoly}[\hatDS_{oo}]^{-1} \eta_o
  = \hatDS_{oo}W_{oo}^{-1} \eta_{o}.
\end{equation}
Since $W_{oo}$ is a matrix close to the identity matrix, the inversion
of $W_{oo}$ is easily performed by any iterative solver within a few
iterations. 
We use the BiCGStab solver in our implementation.

\subsection{Noisy Metropolis test for the correction factor}
\label{sub:NoisyMetropolis}

In order to construct an exact algorithm, we have to take account of
the correction factor $\left(\det[W_{oo}]\right)^2$.
We use the noisy Metropolis test method of Kennedy and
Kuti~\cite{Kennedy_Kuti}, which was previously applied to make the
multiboson algorithm
exact in Refs.~\cite{Borici_Forcrand,Borrelli_Forcrand_Galli}. 

After a trial configuration $U'$ is accepted by the HMC Metropolis
test, we make another Metropolis test for the correction factor.  
Generating a Gaussian vector $\chi_{o}$ with unit variance and zero
mean, the probability $P_{\mathit{corr}}[U\rightarrow U']$ to accept
the trial configuration is given by 
\begin{equation}
  P_{\mathit{corr}}[U\rightarrow U']=\mbox{min}\left[1,e^{-dS}\right],
  \label{eq:NFtwoPHMC_Pcorr}
\end{equation}
where
\begin{equation}
  dS= \left| \left(W_{oo}[U']\right)^{-1}W_{oo}[U]\chi_{o}\right|^2 
     -\left| \chi_{o} \right|^{2}.
\label{eq:NFtwoPHMC_dS}
\end{equation}
The inverse $\left(W_{oo}[U']\right)^{-1}$ is calculated using the
BiCGStab solver as in the generation of the pseudofermion field
$\phi_o$. 

\subsection{Numerical test of the efficiency}
\label{sec:numerical_test_of_efficiency}
The total acceptance rate $P_{\mathit{total}}$ of the PHMC algorithm is
a product of that of the HMC Metropolis test $P_{\mathit{acc}}$ and of
the noisy Metropolis test $P_{\mathit{corr}}$.
In this section we present several numerical tests on how 
$P_{\mathit{acc}}$ and $P_{\mathit{corr}}$ depend on parameters of the
algorithm, such as the order of polynomial, and the MD step size.
The simulations are made on the small heavy lattice, whose
parameters are summarized in Table~\ref{tab:Lattice_Parameters}.

For the choice of polynomial type, 
we test the (non-Hermitian) Chebyshev polynomial
(C-PHMC) and the adapted
polynomial~\cite{Forcrand_UV_adopted,Alexandrou_UV_filter} (A-PHMC). 
The order of the polynomial tested is listed in
Table~\ref{tab:Small_Heavy_Parameters}.

We first study the $\Npoly$ dependence of $P_{\mathit{acc}}$.
In Fig.~\ref{fig:dHvsNpoly_PHMC} we show $\langle dH\rangle$ as a
function of $\Npoly$ on the small heavy lattice at a fixed
$dt=1/32$ ($\Npoly=32$) for both C-PHMC and A-PHMC algorithms. 
We find that $\langle dH\rangle$ is almost independent of $\Npoly$ and
agrees with the same quantity for the S-HMC algorithm.
This is expected if the effective action of the PHMC algorithm
approximates the original action well, because the PHMC 
replaces $[\hatDS_{oo}]^{-1}$ by a polynomial $P_{\Npoly}[\hatDS_{oo}]$
and the two are equivalent if the polynomial is
a good approximation of the inverse.
The MD step size dependence of $\langle dH\rangle$ is plotted in
Fig.~\ref{fig:dHvsdt_PHMC} for the usual S-HMC and for the C-PHMC
with $\Npoly$=26 [which we call C-PHMC(26)], where we find good
agreement among different algorithms.
This means that the acceptance $P_{\mathit{acc}}$ in PHMC is almost the
same as that in the usual HMC.
 
In contrast, $P_{\mathit{corr}}$ is expected to be sensitive to
$\Npoly$.
Since the acceptance rate $P_{\mathit{corr}}$ is directly related to
the expectation value of $dS$ as defined in
Eq.~(\ref{eq:NFtwoPHMC_dS}), we measure the dependence of $\langle dS\rangle$
on $\Npoly$.
Figure~\ref{fig:dSvsNpoly_PHMC} shows the plot for C-PHMC and A-PHMC at
a fixed $dt=1/32$ ($\Npoly=32$). 
The dotted lines represent a fit with an exponential
form~\cite{Borrelli_Forcrand_Galli} 
\begin{equation}
  \label{eq:dS}
  \langle dS\rangle = \pi a^{2}\exp(-2 b \Npoly).
\end{equation}
The exponential form is expected because 
the error of the polynomial approximation behaves
as $|zP_{\Npoly}[z]-1|=(z-1)^{\Npoly+1}$ in the Chebyshev polynomial
case.
We find that the data are well described by the exponential form and
that $\langle dS\rangle$ is much smaller for the adapted polynomial
A-PHMC than that for the Chebyshev polynomial C-PHMC, which demonstrates
the efficiency of the adapted polynomial.

The acceptance rate in the noisy Metropolis step $P_{\mathit{corr}}$ is
related to $\langle dS\rangle$ as 
$P_{\mathit{corr}}=\mbox{erfc}(\langle dS \rangle^{1/2}/2)$.
We then obtain a plot of $P_{\mathit{corr}}$ as a function of
$\Npoly$ in Fig.~\ref{fig:PGMPaccvsNpoly_PHMC}.
The dotted curves represent
\begin{equation}
  \label{eq:Pcorr_vs_Npoly}
  P_{\mathit{corr}} = {\mathrm{erfc}}
  \left(
    \frac{\sqrt{\pi}a}{2} \exp(-b \Npoly)
  \right),
\end{equation}
with $a$ and $b$ the parameters in Eq.~(\ref{eq:dS}).
We clearly see that A-PHMC requires a smaller polynomial order
$\Npoly$ than C-PHMC to achieve the same acceptance rate.
For instance, to obtain $P_{\mathit{corr}}\simeq 0.8$ we need
$\Npoly=24$ for C-PHMC while A-PHMC requires only $\Npoly=18$.

The efficiency of the PHMC algorithm for the noisy Metropolis test step
can be quantified by $P_{\mathit{corr}}/\Npoly$, because the
number of arithmetic operations is roughly proportional to $\Npoly$.
In Fig.~\ref{fig:EffvsPGMPacc_PHMC} we plot 
$P_{\mathit{corr}}/\Npoly$ against $P_{\mathit{corr}}$, and find that
the A-PHMC is about 30\% more efficient than C-PHMC.
We also find that the efficiency peaks around 
$P_{\mathit{corr}}$=0.85.
From Eq.~(\ref{eq:Pcorr_vs_Npoly}) we obtain
\begin{equation}
  \label{eq:Pcorr/Npoly}
  \frac{P_{\mathit{corr}}}{\Npoly} =
  \frac{1}{b}
  \frac{P_{\mathit{corr}}}{
    -\log\left\{
      \frac{1}{a}
      (\sqrt{1-2\log(P_{\mathit{corr}})}-1)
      \right\}}
\end{equation}
using an expansion 
${\mathrm{erfc}}(x)
=\exp\left\{-(2/\sqrt{\pi})[x+(x^2/\sqrt{\pi})]\right\}
+ O(x^3)$.
The efficiency is proportional to $1/b$, which controls the
exponential fall off of $\langle dS\rangle$ as in Eq.~(\ref{eq:dS}), and
the position of the maximum efficiency depends on $a$.
When $a\sim O(10)$, as we observe in these tests, the maximum appears
around $P_{\mathit{corr}}$=0.85 and it moves to larger values of
$P_{\mathit{corr}}$ as $a$ becomes larger.
Since $a$ is expected to scale as
$V^{1/2}(\kappa/\kappa_c)$~\cite{Borrelli_Forcrand_Galli}, the
maximum efficiency is obtained for $P_{\mathit{corr}}>$ 0.85 when the
lattice volume becomes larger or when the sea quark becomes lighter.

\subsection{PHMC on large lattices}
\label{sec:PHMC_on_large_lattices}.

The PHMC algorithm works well with a reasonable order of
the polynomial on the small heavy lattice.  It is not trivial, however, 
whether it really works on larger lattices, because we expect that 
a polynomial with much larger order is needed.

For a numerical test on the ``large heavy'' and ``large light''
lattices (Table~\ref{tab:Lattice_Parameters}) we consider the
Chebyshev polynomial (C-PHMC) only, since we were not able to obtain
an optimized polynomial for the A-PHMC.
The reason is that the minimization of 
$\left|\hatDS_{oo} P_{\Npoly}[\hatDS_{oo}]\eta_{o} - \eta_{o}\right|^{2}$
with respect to the coefficients of polynomial 
failed to converge for polynomials of a large ($\sim 100$) order which 
are needed for these large lattices.
This is likely a problem of the steepest descent algorithm used in
the minimization, and not a fundamental difficulty of the adapted
polynomial.  We leave a resolution of this problem to future studies.

We first consider the question of how the polynomial approximation
of $[\hatDS_{oo}]^{-1}$ works for reasonably large lattices.
To investigate this we define a residual
\begin{equation}
  \frac{
    |P_{\Npoly}[\hatDS_{oo}] \hatDS_{oo}\eta_{o}-\eta_{o}|
    }{|\eta_{o}|},
\label{eq:residualP}
\end{equation}
with a Gaussian noise vector $\eta_{o}$.
We expect that the residual becomes exponentially smaller as $\Npoly$
increases, if the polynomial provides a good approximation of the
inverse Dirac matrix.
We measure this quantity on 20 thermalized configurations of large
heavy and large light lattices and plot it as a function of
$\Npoly$ in Fig.~\ref{fig:Conv_PHMC}.
For both heavy and light dynamical quarks, we find a clear exponential
decrease, while the slope significantly depends on the sea quark mass.
We also note that the polynomial approximation is not distorted by the
round-off error even for $\Npoly\sim$ 100--200. 

When the order of polynomial is large, another important test is the
check of the reversibility in the MD steps.
As we mentioned in Sec.~\ref{subsec:Force_calculation},
our implementation of the force calculation may cause round-off errors.
As in Sec.~\ref{sec:Reversibility} we investigate the violation of 
reversibility in $\langle |\Delta H|/H \rangle$, 
$\langle |\Delta U|\rangle$, and $\langle|\Delta P|\rangle$ by
measuring these quantities on the same 20 configurations.
The results are plotted in Figs.~\ref{fig:DHDUDPrev_LH_PHMC} and
\ref{fig:DHDUDPrev_LL_PHMC}, for large heavy and large light,
as a function of $\Npoly$. 
We observe no dependence on $\Npoly$ for both lattices and the
violation of reversibility remains close to the limit of the double 
precision arithmetic.
This implies that the Clenshaw-type representation of the polynomial 
Eq.~(\ref{eq:PolySumDef})
adopted in our implementation of the PHMC algorithm does not accumulate round-off
errors even for large $N_{poly}$.
We also emphasize that the violation is much smaller than in the usual
HMC plotted in Figs.~\ref{fig:DHDUDPrev_LH_HMC} and
\ref{fig:DHDUDPrev_LL_HMC}.
In the HMC algorithm the number of arithmetic operations can be different
between forward and backward steps, because the convergence of the
BiCGStab solver is controlled by the condition that the
residual is smaller than a certain value.
We suspect that the reversibility becomes better if the number of iterations
(thus the number of arithmetic operations) is fixed in the solver.
Even if this is the case, the numerical stability is not optimized in
the BiCGStab solver, and the PHMC is still expected to perform better 
regarding the reversibility.

We then measure the actual efficiency on large lattices.
The simulation parameters and some results are summarized in 
Tables~\ref{tab:Large_Heavy_Parameters_PHMC} and 
\ref{tab:Large_Light_Parameters_PHMC}
for heavy and light dynamical quarks.
We plot $\langle dS\rangle$ and  
$P_{\mathit{corr}}={\mathrm{erfc}}(\langle dS \rangle^{1/2}/2)$ 
as functions of $\Npoly$ in 
Figs.~\ref{fig:dSvsNpoly_LH_PHMC} and
\ref{fig:PGMPaccvsNpoly_LH_PHMC}.
Compared to the small lattice, substantially larger $\Npoly$ are
needed to keep the acceptance rate at reasonably large values.

Furthermore, $\langle dS\rangle$ and the acceptance depends
substantially on the sea quark mass.
As discussed in Ref.~\cite{Borrelli_Forcrand_Galli} the parameter $b$,
which parametrizes the slope of $\langle dS\rangle$, is expected to be
proportional to the quark mass.
This expectation is confirmed in our simulations:
the ratio of the quark masses in the two simulations is
2.04(6), while the ratio of $b$ is 2.15(15).

The efficiency of the noisy Metropolis step $P_{\mathit{corr}}/\Npoly$
is plotted in Fig.~\ref{fig:EffvsPGMPacc_LH_PHMC}.
The maximum efficiency is achieved around $P_{\mathit{corr}}=0.9$, and
the height at the maximum is lower for the lighter quark mass
than that for the heavier one by about a factor of two,
as we expected from the ratio of $b$
[and from Eq.~(\ref{eq:Pcorr/Npoly})].

Finally, we compare the total efficiency of the PHMC algorithm with
that of the usual HMC. 
The efficiency is parametrized as
$P_{\mathit{total}}/[\NMult/\mathrm{traj}]$,
which is plotted in Fig.~\ref{fig:PtotaccovMult_LH_PHMC} against
the total acceptance ratio $P_{\mathit{total}}$.
The total acceptance ratio $P_{\mathit{total}}$ of the PHMC algorithm is
defined by $P_{\mathit{total}}=P_{\mathit{acc}}P_{\mathit{corr}}$;
for the HMC algorithm it is $P_{\mathit{total}}=P_{\mathit{acc}}$.
The number of hopping matrix multiplications to cover a unit trajectory,
$[\NMult/\mathrm{traj}]$,
is counted in the program.
The efficiency of PHMC is slightly better than the usual HMC for both
heavy and light dynamical quarks.
We note that the efficiency of HMC depends substantially on
the stopping condition imposed.
As we discussed in Sec.~\ref{sec:Reversibility}, we carefully chose
the stopping condition for HMC, but the remaining violation of the
reversibility is still large compared to the PHMC.
Therefore, in order to guarantee the exactness of the algorithm strictly,
a strict stopping condition is required and then the efficiency of
HMC becomes much lower.

\section{PHMC algorithm for an odd number of flavors}
\label{sec:OddFlavor}

In this section we describe an extension of the PHMC algorithm to the
case of odd number of flavors.
As we already outlined in Sec.~\ref{sec:Outline_of_the_algorithm}, the
algorithm is almost the same as that for even number of flavors,
except for the polynomial in the evaluation of the correction factor. 
We make a numerical check of the algorithm by comparing the simulation
of $1+1$-flavor QCD with the two-flavor case simulated with the
HMC and PHMC algorithms.
In addition we carry out a simulation of $2+1$-flavor QCD and compare
the results with that obtained by the $R$ algorithm.

\subsection{PHMC for one-flavor QCD}

In order to construct a real and positive definite effective action
for one-flavor of dynamical quark, we use the trick proposed by 
Bori\c{c}i and de~Forcrand~\cite{Borici_Forcrand} and 
Alexandrou \textit{ et al.}~\cite{Nf_one_Alexandrou}, which was already
described in Sec.~\ref{sec:outline_odd_flavor}.

A polynomial of even degree $P_{\Npoly}[z]$ can be split into 
the product of two
polynomials $T_{\Npoly}[z]$ and $\overline{T}_{\Npoly}[z]$ as
\begin{eqnarray}
  P_{\Npoly}[z] &=& T_{\Npoly}[z] \overline{T}_{\Npoly}[z], \\
  \label{eq:PTT}
  T_{\Npoly}[z] & \equiv & \sum_{i=0}^{\Npoly/2} d_{i} (z-1)^{i}, 
  \label{eq:TP}  \\
  \overline{T}_{\Npoly}[z] & \equiv & \sum_{i=0}^{\Npoly/2} d^{*}_{i} (z-1)^{i}.
  \label{eq:TprimeP}
\end{eqnarray}
Note that here we use the summation representation for $T_{\Npoly}[z]$ 
($\overline{T}_{\Npoly}[z]$) instead of the product representation as in 
Eq.~(\ref{eq:SplitdetD}).
The coefficients $d_{i}$ in $T_{\Npoly}[z]$ are determined as follows.
First, we consider the product representation of $P_{\Npoly}[z]$ as
$P_{\Npoly}[z]=c_{\Npoly}\prod_{k=1}^{\Npoly}(z-z_k)$.
The ordering of the monomials is defined so that $\arg(z_{k}-1)$ 
increases monotonically with increasing $k$.
Since the roots $z_{k}$ appear with their complex conjugate,
 we find $z_{k}=z_{\Npoly+1-k}^{*}$ $(k=1\ldots\Npoly/2)$.
We then split the polynomial into the product of two polynomial
as $P_{\Npoly}[z]=c_{\Npoly}\prod_{j=1}^{\Npoly/2}(z-z_{k(j)})(z-z_{k(j)}^{*})$,
where the reordering index $k(j)$ is defined by $k(j)=2j-1$.
Then we obtain a ``square root'' of the polynomial as
$T_{\Npoly}[z]=\sqrt{c_{\Npoly}} \prod_{j=1}^{\Npoly/2} (z-z_{k(j)})$, from
which we arrive at the polynomial representation Eq.~(\ref{eq:TP}) by expanding
the product representation.
Since we do not use the product representation of $T_{\Npoly}[z]$ in
the numerical simulation, the problem of the ordering of monomials is
irrelevant as long as one uses long enough decimal precision or
computer algebra systems to obtain the coefficients $d_i$.

We note that $T_{\Npoly}[z]^\dagger\not=\overline{T}_{\Npoly}[z]$
for complex $z$, but for the determinant of the Wilson-Dirac operator $D$ one
can prove the relation
\begin{equation}
  \label{eq:trick}
  \det[T_{\Npoly}[D]]^* = \det[\overline{T}_{\Npoly}[D]],
\end{equation}
using the $\gamma_5$ Hermiticity property 
$D^\dagger=\gamma_5 D\gamma_5$.  It follows that
\begin{equation}
\det[P_{\Npoly}[D]] 
= \det[\overline{T}_{\Npoly}[D]] \cdot \det[T_{\Npoly}[D]]
= |\det[T_{\Npoly}[D]]|^2.
\end{equation}
For the preconditioned case, the Hermiticity is modified to
$\left.\hatDS_{oo}\right.^{\dag}=
\gamma_{5}(1+T)_{oo}\hatDS_{oo}(1+T)_{oo}^{-1} \gamma_{5}$, for which
Eq.~(\ref{eq:trick}) holds as well.

The partition function for one-flavor QCD can be written as
\begin{eqnarray}
  {\cal Z}
  & = &
  \int\!\!{\cal D}U{\cal D}P{\cal D}\phi_o^{\dag}{\cal D}\phi_o\,
  \det[W_{oo}] e^{-H_{\mathrm{PHMC}}[P,U,\phi_{o}]}, 
  \nonumber \\
  H_{\mathrm{PHMC}}[P,U,\phi_{o}]
  & = &
  \frac{1}{2}P^2 
  + S_{g}[U]  
  + S^{S}_{\mathit{poly}}[\phi_o] 
  + S^{S}_{\mathit{det}}[U], 
  \nonumber \\
  S^{S}_{\mathit{poly}}[\phi_o]&=&
  \left|T_{\Npoly}[\hatDS_{oo}]\phi_{o}]\right|^{2},
  \nonumber \\
  S^{S}_{\mathit{det}}[U]
  & = &
  -\left(\log\det[1+T_{ee}]+\log\det[1+T_{oo}]\right).
  \label{eq:partition_function_H_odd}
\end{eqnarray}
The polynomial $P_{\Npoly}[\hatDS_{oo}]$ in the
two-flavor case Eq.~(\ref{eq:partition_function_H_PHMC}) is 
replaced by $T_{\Npoly}[\hatDS_{oo}]$.
The correction factor $\det[W_{oo}]$ is the same as that
defined in Eq.~(\ref{eq:partition_function_H_PHMC}),
but the exponent is 1.

Every step of the HMC part of the simulation is the same as the
corresponding step in the two-flavor case, except that we use the
polynomial $T_{\Npoly}$ rather than $P_{\Npoly}$.
The pseudofermion field is similarly generated by
\begin{equation}
  \label{eq:PS_generation_Nf_one}
  \phi_{o}=
   T_{\Npoly}[\hatDS_{oo}]^{-1}\eta_{o}=
   \overline{T}_{\Npoly}[\hatDS_{oo}]\hatDS_{oo}W_{oo}^{-1}\eta_{o},
\end{equation}
with a Gaussian noise vector $\eta_{o}$ at the beginning of each MD step.
On the other hand, the noisy Metropolis step to incorporate the
correction factor requires a special treatment, because the correction
factor is not $\det[W_{oo}]^2$ but $\det[W_{oo}]$.

\subsection{Noisy Metropolis test for the one-flavor case}
\label{sub:NoisyMetropolisOddFlavor}. 

If the fermion determinant $\det[\hatDS_{oo}]$ is positive,
$\det[W_{oo}]$ is also positive and its square root is well defined. 
We calculate the square root of the matrix $W_{oo}$ by solving the
equation $A_{oo}^2=W_{oo}$ using the Taylor expansion
\begin{equation}
  \label{eq:A_oo}
  A_{oo} = 
  1 + \sum_k^\infty \frac{(2k-3)!!}{(2k)!!} \delta_{oo}^k
  = 
  1 + \frac{1}{2}\delta_{oo} - \frac{1}{8}\delta_{oo}^2 + 
  \frac{1}{16}\delta_{oo}^3 \cdots
\end{equation}
with $\delta_{oo}\equiv W_{oo}-1$, because we expect that $W_{oo}$ is
close to the identity matrix when the polynomial
$P_{\Npoly}[\hatDS_{oo}]$ is a good approximation of
$(\hatDS_{oo})^{-1}$.
We obtain
\begin{equation}
  \label{eq:W=A^2}
  \det[W_{oo}] = \left|\det[A_{oo}]\right|^{2},
\end{equation}
using the (preconditioned) $\gamma_5$ Hermiticity property 
$A_{oo}^\dagger=\gamma_5(1+T)_{oo} A_{oo}(1+T)^{-1}_{oo}\gamma_5$.

Once we obtain the matrix $A_{oo}$, we can perform the noisy
Metropolis test Eq.~(\ref{eq:NFtwoPHMC_Pcorr}) replacing $W_{oo}$ in
Eq.~(\ref{eq:NFtwoPHMC_dS}) by $A_{oo}$,
\begin{equation}
  dS= \left| \left(A_{oo}[U']\right)^{-1}A_{oo}[U]\chi_{o}\right|^2 
     -\left| \chi_{o} \right|^{2}.
     \label{eq:NFonePHMC_dS}
\end{equation}
The only complication is the use of the Taylor expansion
Eq.~(\ref{eq:A_oo}) every time we need a multiplication with $A_{oo}$.
For the inverse $A_{oo}^{-1}$ we use another polynomial
\begin{equation}
  \label{eq:Ainv_oo}
  A_{oo}^{-1} = 
  1 + \sum_k^\infty (-1)^k \frac{(2k-1)!!}{(2k)!!} \delta_{oo}^k
  = 
  1 - \frac{1}{2}\delta_{oo} + \frac{3}{8}\delta_{oo}^2 -
  \frac{5}{16}\delta_{oo}^3 \cdots.
\end{equation}
In the numerical calculation, summation from the lower order to the
higher should be avoided to reduce round-off errors.
We therefore use the following (Clenshaw's type) expressions:
\begin{eqnarray}
 A_{oo} & = &
 \left[ 1 + \frac{ 1}{2}\delta_{oo} 
   \left[ 1 + \frac{-1}{4}\delta_{oo} 
     \left[ 1 + \frac{-3}{6}\delta_{oo} 
       \cdots 
       \left[ 1 + \frac{3-2 k}{2 k}
         \delta_{oo} \right]
       \cdots 
     \right]
   \right]
 \right],
 \\
 {A'}^{-1}_{oo} & = &
 \left[ 1 + \frac{-1}{2}\delta'_{oo} 
   \left[ 1 + \frac{-3}{4}\delta'_{oo} 
     \left[ 1 + \frac{-5}{6}\delta'_{oo} 
       \cdots 
       \left[ 1 + \frac{1-2 k}{2 k}
         \delta'_{oo} \right]
       \cdots 
     \right]
   \right]
 \right].
\end{eqnarray}
A shortcoming of this method is that we have to recalculate the entire 
expressions when we need to increase the order of truncation $k$ in
the Taylor expansion. 

In order to avoid systematic errors from the truncation of the Taylor
expansion, we monitor the residual 
\begin{equation}
  r_{1} = 
  \frac{\left|A_{oo}[U](A_{oo}[U]\chi_{o}) - W_{oo}[U]\chi_{o}\right|}{
    |W_{oo}[U]\chi_{o}|},
\label{eq:Res1}
\end{equation}
in the calculation of $A_{oo}[U]\chi_{o}$, and
\begin{equation}
  r_{2} = 
  \frac{\left|W_{oo}[U'](A_{oo}[U'])^{-1}
              \left\{(A_{oo}[U'])^{-1}\omega_{o}\right\}
      -\omega_{o}\right|}{|\omega_{o}|},
\label{eq:Res2}
\end{equation}
in the calculation of $(A_{oo}[U'])^{-1}\omega_{o}$ with
$\omega_{o}=A_{oo}[U]\chi_{o}$.
We require that the residuals be smaller than $10^{-14}$ to keep
the exactness of the algorithm. 
In the simulation program we always monitor the residuals, and when
the residuals become larger than our condition we repeat the
calculation increasing $k$ until it becomes satisfied.

The necessary order of the Taylor expansion depends significantly on
the order of polynomial $\Npoly$.
If $\Npoly$ is large enough, $W_{oo}$ is very close to the identity and the
Taylor expansion may be truncated at very low orders.
Therefore, there is a complicated trade-off between $\Npoly$ and $k$
to the computational cost in the algorithm. 
We consider briefly the computational cost to calculate the square root of the
correction matrix and the noisy Metropolis acceptance probability as follows.  
In the case of the Chebyshev polynomial, the residual
of the correction matrix is estimated as
\begin{equation}
  W_{oo} - 1 = \delta_{oo} = (\hatDS_{oo} - 1)^{\Npoly+1}.
\end{equation}
If we take $\lambda$ as the largest eigenvalue of $\hatDS_{oo} - 1$, 
this leads to 
$\left|\delta_{oo}\right| \simeq \left|\lambda\right|^{\Npoly+1}$
where $\left|\lambda\right| <1$ is assumed.
To keep the residual of the square root Eq.~(\ref{eq:Res1}) (for example)
lower than a constant $\epsilon$, we have the following inequality when the
Taylor expansion is truncated at an order $k$:
\begin{equation}
  r_{1}\propto \left|\delta_{oo}^{k+1}\right| = \left|\lambda\right|^{(k+1)(\Npoly+1)}
< C \epsilon,
\label{eq:IneqCost}
\end{equation}
with a coefficient $C$.
Thus $(k+1)(\Npoly+1)$ must be larger than a constant proportional to $\ln(\epsilon)$. 
When we fix $\epsilon$ as a stopping condition, the truncation order $k$ is 
chosen so as to satisfy Eq.~(\ref{eq:IneqCost}).
The computational cost to calculate the square root of the correction matrix
becomes a constant because the number of multiplication of 
$\hatDS_{oo}$ is proportional 
to $k\times \Npoly$, which is roughly $\sim (k+1)(\Npoly+1)$.
Consequently the total amount of the computational cost to calculate 
Eq.~(\ref{eq:NFonePHMC_dS}) becomes almost constant.
Thus we conclude that the choice of $\Npoly$ does not affect the cost in the 
noisy Metropolis test, and that the efficiency of the whole 
algorithm is governed by the cost of the molecular dynamics step 
(proportional to $\Npoly$) and  
by the acceptance rates of the HMC and the noisy Metropolis tests.

In order to evaluate the correction factor $\det[W_{oo}]$,
Takaishi and de~Forcrand~\cite{Takaishi_Forcrand_Nf3} 
employed the idea
of the unbiased stochastic estimator~\cite{Lin_Liu_Sloan} using
$\sqrt{\det[W_{oo}[U']^2/W_{oo}[U]^2]}
=\sqrt{\langle  e^{-dS}\rangle_{\chi_o}}$ 
from several estimates of
$\langle e^{-dS}\rangle_{\chi_o}$ with $dS$ defined in 
Eq.~(\ref{eq:NFtwoPHMC_dS}) for the $N_f=2$.
Their method is faster than ours because they do not need to
calculate the square root of the correction matrix as we did in 
Eqs.~(\ref{eq:A_oo}) and (\ref{eq:Ainv_oo}).
On the other hand, the stochastic estimator may produce
negative probabilities for the Metropolis test, which leads to 
systematic errors in the final results.
In order to avoid this problem they keep $dS$ sufficiently small
with a high acceptance ratio so that the negative probabilities
within a desired trajectory length do not appear.
In our method these problems are avoided at the price of additional
computational costs by taking explicitly the square root of the
correction matrix.

\subsection{Numerical test with $(1+1)$-flavor QCD}

The algorithm for one-flavor of dynamical fermion can be tested by
considering $(1+1)$-flavor QCD, which should be identical to two-flavor
QCD.
Since we already have results with established algorithms for
two-flavor QCD, we check if we can reproduce the results with the
$(1+1)$-flavor QCD simulation. 
For $(1+1)$ flavors, we introduce two sets of pseudofermion fields
$\phi_o^{[f]}$ ($f$=1, 2) with the effective action 
$S_{\mathit{poly}}^S[\phi_o^{[f]}]
= |T_{\Npoly}[\hatDS_{oo}]\phi_o^{[f]}|^2$.
The correction factor $\det[W_{oo}]$ is evaluated twice with the noisy
Metropolis test described in Sec.~\ref{sub:NoisyMetropolisOddFlavor}.

Simulation parameters and some results on our small heavy lattice
are listed in Table~\ref{tab:CompPlaqNf1plus1}.
We employ the Chebyshev polynomial of order $\Npoly=26$ both in the
two-flavor simulation and in the $(1+1)$-flavor simulation with PHMC algorithms.
Note that the order of the polynomial $T_{\Npoly}$ 
in $N_f=1+1$ is $26/2=13$ by its definition for each pseudofermion.
We also have a result with the standard S-HMC.

We observe in Table~\ref{tab:CompPlaqNf1plus1} that the 
three algorithms give a consistent plaquette expectation value within
the statistical error of less than 0.1\%.  It is evident that
the algorithm for odd number of flavors works as we expected.
The statistical error is evaluated with the binned jack-knife method and
the bin size is increased until the error ceases to grow.

In the same table we find that $\langle dH\rangle$, which controls 
the HMC acceptance $P_{\mathit{acc}}$, is significantly
smaller for the $(1+1)$-flavor simulation at the same MD step size $dt$.
The size of $\langle dH\rangle$ depends on the
precise form of the Hamiltonian we consider. 
While the formula described in Ref.~\cite{Gupta_Kilcup_Sharpe_88} may be 
employed to examine this issue, we do not pursue it here 
because of the complication of the force contribution from the 
pseudofermion action.   Note that this decrease of $\langle dH\rangle$ in 
the $N_f=1+1$ case does not immediately mean
an increase of the efficiency.  The reason is that we expect the duplication 
of the pseudofermion field to cause an extension of the autocorrelation time. 

We find that the acceptance rate $P_{\mathit{corr}}^{N_f=2}$ in
the correction factor for the two-flavor case is related to those of the
$(1+1)$-flavor simulation as
$P_{\mathit{corr}}^{N_f=2}\simeq (P_{\mathit{corr}}^{N_f=1})^2$. 
This property can be explained as follows:
Expanding $dS^{N_f=2}$ in Eq.~(\ref{eq:NFtwoPHMC_dS}) in terms of 
$\delta_{oo}=W_{oo}[U]-1$ and $\delta'_{oo}=W_{oo}[U']-1$, we obtain 
$dS^{N_{f}=2}
=2\mathrm{Re}[\chi^{\dag}_{o}(\delta-\delta')_{oo}\chi_{o}]$
up to $O(\delta^2,\delta'^{2},\delta \delta')$. 
On the other hand, $dS^{N_{f}=1}$ in Eq.~(\ref{eq:NFonePHMC_dS}) is expressed as 
$dS^{N_{f}=1}= 
\mathrm{Re}[\chi^{\dag}_{o}(\delta-\delta')_{oo}\chi_{o}]$.
Up to higher orders in $\delta_{oo}$ and $\delta_{oo}'$ we then obtain
$dS^{N_f=2}\simeq 2 dS^{N_f=1}$ and 
$P_{\mathit{corr}}^{N_f=2} \simeq (P_{\mathit{corr}}^{N_f=1})^2$.

We also test our algorithm on large heavy and large light lattices.
The convergence of the polynomial 
$T_{\Npoly}[\hatDS_{oo}]$ and of the Taylor expansion of the correction factor
is non-trivial on these large lattice sizes. 
To investigate the convergence of 
the polynomial $T_{\Npoly}[\hatDS_{oo}]$ 
we perform the same check as that made for $P_{\Npoly}[\hatDS_{oo}]$.
In Fig.~\ref{fig:Conv_PHMC_Nf1plus1} we show the convergence behavior
using
\begin{equation}
  \frac{\left|
      \overline{T}_{\Npoly}[\hatDS_{oo}]
      T_{\Npoly}[\hatDS_{oo}]\hatDS_{oo}\eta_{o}-\eta_{o}\right|}{|\eta_{o}|},
  \label{eq:residualTT}
\end{equation}
as the residual.
Here $\eta_{o}$ is a Gaussian noise vector and the measurement
is made on 20 thermalized configurations separated by ten trajectories.
Since $\overline{T}_{\Npoly}[\hatDS_{oo}]T_{\Npoly}[\hatDS_{oo}]$
should be $P_{\Npoly}[\hatDS_{oo}]$ by definition,
Eq.~(\ref{eq:residualTT}) must be identical to Eq.~(\ref{eq:residualP})
except for round-off errors.
As shown in Fig.~\ref{fig:Conv_PHMC_Nf1plus1}, Eq.~(\ref{eq:residualTT}) 
decreases exponentially as $N_{\Npoly}$ increases, 
which is the same behavior as
in Fig.~\ref{fig:Conv_PHMC}.
Thus we confirm that there is no unexpected accumulation of round-off errors in
the calculation of $T_{\Npoly}[\hatDS_{oo}]$ with our choice of
$\Npoly$ ($T_{\Npoly}[\hatDS_{oo}]$ is also evaluated with 
the Clenshaw's recurrence formula).
The violation of reversibility is extremely small as plotted in
Figs.~\ref{fig:DHDUDPrev_LH_PHMC_Nf1plus1} and  
\ref{fig:DHDUDPrev_LL_PHMC_Nf1plus1}.
Their magnitude stays around the limit of the double precision
arithmetic, which parallels our finding with the two-flavor case
(Figs.~\ref{fig:DHDUDPrev_LH_PHMC} and \ref{fig:DHDUDPrev_LL_PHMC}). 

Figures~\ref{fig:dSconv_LH_PHMC_Nf1plus1} (large heavy)
and \ref{fig:dSconv_LL_PHMC_Nf1plus1} (large light)
show the convergence behavior of the Taylor expansion of 
the correction matrix as a function of the order of the expansion.
The convergence is monitored with the residuals $r_1$ and $r_2$
defined in Eqs.~(\ref{eq:Res1}) and (\ref{eq:Res2}), respectively.
We also monitor the convergence of the weight $dS$ defined in Eq.~(\ref{eq:NFonePHMC_dS}),
by measuring $|dS-dS_{\mathit{end}}|$,
where $dS_{\mathit{end}}$ is the value of $dS$ at the highest order of 
the expansion.
These figures are also plotted with measurements on 20 configurations
separated by 10 trajectories. 
Open symbols are obtained for the smallest $\Npoly$ 
(70 for large heavy, 100 for large light), 
and filled ones are for the largest $\Npoly$
(190 for large heavy, 200 for large light).
The convergence of the residuals is almost exponential.
The slope, however, becomes weaker near the limit of the double
precision arithmetic. 
In the region where the exponential decay is observed, $k\times \Npoly$ seems
to behave as roughly constant irrespective of the choice of $\Npoly$. This is the expected behavior
discussed in Sec.~\ref{sub:NoisyMetropolisOddFlavor}.
When the stopping condition for $r_1$ and $r_2$ is set to be $10^{-14}$,
the improvement of $|dS-dS_{\mathit{end}}|$ stops at $\sim 10^{-12}$.
Since $dS$ itself is of $O(10^{-2})$, we expect that $dS$ has $\sim 10$ 
digits of significant figure, 
which we expect to be sufficient for current simulation trajectory lengths.
The negative eigenvalue problem did not occur in these investigations,
probably because of the intermediate quark mass we employed.

Table~\ref{tab:CompPHMCNf1plus1_Large} shows the simulation
statistics for C-PHMC with $N_f=1+1$ on both of the large lattices.
We obtain results for the averaged plaquette value which are consistent  
with those for the $N_f=2$ case.  The relation 
$P_{\mathit{corr}}^{N_f=2} \simeq (P_{\mathit{corr}}^{N_f=1})^2$ holds
again for such large lattice sizes, and we did not encounter the negative 
eigenvalue problem during the long trajectories $(\sim1000)$.
We expect that the total efficiency has the same functional dependence 
on $\Npoly$ as that with the $N_f=2$ PHMC, since 
the behavior on $\Npoly$ is mostly ruled by the molecular dynamics.
The actual value of the total efficiency is slightly worse
than that with the $N_f=2$ PHMC algorithm due to the two pseudofermion
generations, the Hamiltonian calculation,
and monitoring of the residual in the noisy Metropolis test.
We note that the autocorrelation time may be extended by the increase 
of the dynamical variable in the path integral. Examination of this 
point is left for future studies.
With the numerical tests described here we conclude that our PHMC algorithm 
for one-flavor dynamical quark works well even for a moderately large
lattice size $16^3\times 48$ at intermediate quark masses of 
$m_{\mathit{PS}}/m_{\mathit{V}}\sim 0.7-0.8$, at least in the
$N_f=1+1$ case. 

\subsection{A $(2+1)$-flavor QCD simulation}

Combining the two-flavor HMC algorithm with the one-flavor PHMC
leads to an exact algorithm for $(2+1)$-flavor QCD.
For the two-flavor part, we may also choose the PHMC if the usable 
amount of memory allows to store work vectors.
A test of the algorithm can be performed comparing the results with
those of the $R$ algorithm~\cite{Gottlieb_etal_Hybrid_R}
after an extrapolation to zero step size in the latter.
In this section we show the results of such a comparison 
on some small lattices. 

The numerical test is made with the following two sets of lattice parameters.
One set uses a lattice of size $4^3\times 8$ at $\beta=4.8$, sea quark mass of 
$\kappa_{ud}=0.150$ for two light flavors, and $\kappa_{s}=0.140$ for
the third flavor, and $c_{\mathrm{sw}}=1.0$ for all three flavors
($N_f=2+1$ small).  The order of the polynomial is set to $\Npoly=10$ for the
single flavor.
The second set uses a $8^3\times 16$ lattice, $\beta=5.0$, 
$\kappa_{ud}=0.1338$, and $\kappa_{s}=0.1330$, and $c_{\mathrm{sw}}=2.08$
($N_f=2+1$ middle), where $\Npoly=58$ is employed.
For both lattice sizes we use the Chebyshev polynomial with unit circle
convergence domain.

The simulation statistics is tabulated in
Tables~\ref{tab:HR_Comp_Small} and \ref{tab:HR_Comp_Middle} together
with the plaquette expectation value extracted from the $R$ algorithm.
Figures~\ref{fig:HR_Comp_Small} and \ref{fig:HR_Comp_Middle} 
show the plaquette expectation value from the runs with the $R$ algorithm 
at several values of the MD step size $dt$ (open symbols).
Filled symbols are from the PHMC algorithm.
The plaquette values with the $R$ algorithm extrapolated to zero step size
are plotted with dotted horizontal lines.
We observe that our exact algorithm (filled symbols) gives 
results at a finite $dt$ 
(see also Tables~\ref{tab:HR_Comp_Small} and \ref{tab:HR_Comp_Middle})
consistent with the extrapolated value (horizontal dotted line)
of the $R$ algorithm.
Because of the finite $dt$ dependence, the cost to obtain reliable
results with the $R$ algorithm is higher than that of the PHMC
algorithm. 

For larger and realistic lattice sizes, we started a parameter search
in order to realize a physical volume $L\sim 1.7-2.0$ fm, 
a lattice cutoff $a^{-1}\sim 1.5-2.0$ GeV, and pseudoscalar to
vector meson mass ratios
$m_{\mathit{PS}}/m_{\mathit{V}}\sim 0.7-0.8$. During the parameter search we 
found an unexpected first-order phase transition~\cite{JLQCD_1st_order}. 
Details of this search, including the property of the PHMC algorithm with the 
realistic parameters in the $N_f=2+1$ case on large lattice sizes, will be 
reported elsewhere.

\section{Conclusions}
\label{sec:Conclusion}
In this paper, we introduced a polynomial hybrid Monte Carlo (PHMC) algorithm 
which is applicable to QCD with an odd number of flavors.
The algorithm is an extension of the one by Takaishi and 
de~Forcrand~\cite{Takaishi_Forcrand_Nf3} to the $O(a)$-improved
Wilson quark action. We also described a method to remove the systematic error
from the non-Hermitian polynomial approximation to the 
invese of the Wilson-Dirac operator in the single flavor case.

An important technical point uncovered in our work concerns 
the choice of the even-odd preconditioning 
to the $O(a)$-improved Wilson-Dirac operator. 
Asymmetric and symmetric even-odd preconditionings were introduced and 
investigated in the HMC algorithm with two-flavor dynamical quarks.
We found that the HMC algorithm with the {\it symmetrically} even-odd 
preconditioned form of the lattice QCD partition function yields roughly 
a factor two gain in efficiency over the unpreconditioned one.  
This performance exceeds the gain of about 1.5 for the asymmetrical 
preconditioning employed in actual simulations so far.
We, then, decided to use the symmetrically even-odd preconditioned form for 
the quark determinant for the PHMC algorithm.

We explored distinctive features of the PHMC algorithm using the case of 
two flavors of quarks where comparisons with the standard HMC are possible.
Our findings are 
(i) the reversibility is much better with the PHMC algorithm because of the 
fully deterministic nature of multiplication with the Wilson-Dirac operator 
in the force calculation in the molecular dynamics step, 
(ii) for the order of the polynomial chosen sufficiently large, the total 
efficiency of the PHMC algorithm is almost identical to or rather better than
that with the HMC algorithm.  Hence the PHMC algorithm is an alternative 
for $N_f=2$ dynamical QCD simulations on moderately large lattice size
in the intermediate quark mass region 
$m_{\mathit{PS}}/m_{\mathit{V}}\sim 0.7$-$0.8$.

We demonstrated the consistency and applicability of the PHMC algorithm for 
an odd number of flavors by considering the case of two single-flavor 
pseudofermions ($N_f=1+1$ QCD) and comparing it with the established
algorithm for the two-flavor pseudofermion ($N_f=2$ QCD).
The reversibility holds to almost the same degree as that with 
the $N_f=2$ PHMC algorithm.
The noisy Metropolis test for single-flavor part, in which we have to take 
the square root of the correction matrix explicitly, works well on 
moderately large lattices with intermediate quark masses of 
$m_{\mathit{PS}}/m_{\mathit{V}}\sim 0.7-0.8$.

Finally we constructed a PHMC algorithm for $2+1$ flavors of quarks by 
combining a two-flavored pseudofermion, which is employed in the usual 
HMC algorithm, and a single-flavored pseudofermion described by the 
polynomial approximation.
Running the algorithm on two small lattice sizes we confirmed an agreement 
of plaquette values with those from the $R$ algorithm after an extrapolation 
to the zero step size in the latter.  

We conclude that the PHMC algorithm is a viable choice for
realistic simulations of lattice QCD with 2+1 flavors.
Since our numerical tests show that the computational cost
for two single-flavor pseudofermions is comparable to that
of the two-flavor case, the cost for the single-flavor part of
the $(2+1)$-flavor QCD is about a half of the two-flavor part.
We thus expect that the simulation of the $(2+1)$-flavor QCD may
be performed with a cost of a factor $1.5-2$ compared to the
two-flavor QCD simulation.

\begin{acknowledgments}
This work is supported by the Supercomputer Project No.66 (FY2001) of High
Energy Accelerator Research Organization (KEK), and also in part by the
Grant-in-Aid of the Ministry of Education (Nos. 10640246, 11640294, 12014202,
12640253, 12640279, 12740133, 13640260, and 13740169). K-I.I. and N.Y. are
supported by the JSPS.
\end{acknowledgments}

\clearpage
\appendix

%%%%%% Force
\section{Force calculation in the HMC algorithms}
\label{sec:Appendix_A}

In this appendix we describe the explicit form of the quark force in
the HMC algorithms for different preconditionings.
Since most of the definitions and extractions of the quark force are common to
the standard Wilson quark action, we only show the variation of the quark action
under an infinitesimal change of the gauge link variable
as defined in Eq.~(\ref{eq:ForceDef}).

\subsection{Without preconditioning}

If we do not apply the even-odd preconditioning, 
the force from the pseudofermion field is simply written as
\begin{equation}
  \delta S_{q}=\left\{ - X^{\dag} \delta D Y \right\} + \mbox{H.c.},
\label{NonpreconditionedForce}
\end{equation}
where
\begin{eqnarray}
  X&=& \left(D^{\dag}\right)^{-1} D^{-1}\phi, \\
  Y&=& D^{-1}\phi, \\
  \delta D&=& \left(
    \begin{array}{cc}
      \delta T_{ee} & \delta M_{eo} \\
      \delta M_{oe} & \delta T_{oo} 
    \end{array} 
  \right).
\end{eqnarray}
The contribution from the derivative of the hopping matrix 
$\delta M_{eo (oe)}$ is the same as that in the Wilson action.
The contribution from the SW term $\delta T_{ee (oo)}$ is shown in  
Fig.~\ref{fig:ForceDet}, where {\Large $\times$} is a $3\times 3$
matrix defined by
\begin{equation}
\left(\mbox{\Large$\times$}\right)_{\mu\nu}(n)=\left\{-\frac{i
c_{\mathrm{sw}}\kappa}{8}\mbox{tr}_{\mathrm{dirac}}\left[\sigma_{\mu\nu}
Y(n)X(n)^{\dag}\right]\right\}
+\mbox{H.c.}
\end{equation}
$\mbox{tr}_{\mathrm{dirac}}[\cdots]$ means the trace over the spinor
indices. 

\subsection{Asymmetric preconditioning}
The force from the pseudofermion field with the asymmetric
preconditioning is given by
\begin{equation}
  \delta S^{A}_{q}= \left\{-X^{A \dag} \delta D Y^{A}\right\}
  + \mbox{H.c.},
  \label{eq:AsymmetricForce}
\end{equation}
where
\begin{eqnarray}
X^{A} &=& \left(
                 \begin{array}{c}
                     -(1+T)_{ee}^{-1} {M_{oe}}^{\dag} \hat{X}^{A}_{o} \\
                      \hat{X}^{A}_{o}
                 \end{array}
          \right), \\
Y^{A} &=& \left(
                  \begin{array}{c}
                     -(1+T)_{ee}^{-1} M_{eo} \hat{Y}^{A}_{o} \\
                      \hat{Y}^{A}_{o}
                  \end{array}
                  \right), \\
\hat{X}^{A}_{o} &=&  \left(\left.\hatDA_{oo}\right.^{\dag}\right)^{-1}
                     \left(\hatDA_{oo}\right)^{-1} \phi_{o},\\
\hat{Y}^{A}_{o} &=&  \left(\hatDA_{oo}\right)^{-1} \phi_{o}.
\end{eqnarray}
Note that 
$\left.\hatDA_{oo}\right.^{\dag}=\gamma_{5}\hatDA_{oo}\gamma_{5}$
 and $M_{oe}^{\dag}=\gamma_{5}M_{eo}\gamma_{5}$. 

In addition we need the force from the determinant of the SW term,
\begin{equation}
\delta S^{A}_{\mathit{det}}=-2\mbox{Tr}\left[\delta
                      T_{ee}(1+T)^{-1}_{ee}\right].
\end{equation}
This is calculated only for even sites.
The term {\Large$\times$} in Fig.~\ref{fig:ForceDet} from the SW
term is replaced by 
\begin{eqnarray}
  \left(\mbox{\Large$\times$}\right)_{\mu\nu}(n)&=&\left\{
    -\frac{i c_{\mathrm{sw}}\kappa}{8}
    \mbox{tr}_{\mathrm{dirac}}\left[\sigma_{\mu\nu}Y^{A}(n)X^{A}(n)^{\dag}\right]
  \right. \nonumber \\
  &&\left.
    -\frac{i c_{\mathrm{sw}}\kappa}{8}
    \mbox{tr}_{\mathrm{dirac}}\left[\sigma_{\mu\nu}(1+T)^{-1}(n)\right]\delta_{n,{\mathrm{even site}}}\right\}
  \nonumber \\
  &&+\mbox{H.c.}
\end{eqnarray}

\subsection{Symmetric preconditioning}
For the symmetric preconditioning the force is separated into two
parts as
\begin{equation}
\delta S^{S}_{q}= \left\{
                    -X^{S \dag} \delta M Y^{S}
                    -X^{S \dag} \delta T Z^{S}\right\}
                    + \mbox{H.c.},
\label{eq:SHMCaction}
\end{equation}
where
\begin{eqnarray}
\delta M&=& \left(
            \begin{array}{cc}
               0 & \delta M_{eo} \\
               \delta M_{oe} & 0
            \end{array} 
            \right), \\
\delta T&=& \left(
            \begin{array}{cc}
                \delta T_{ee} & 0 \\
                0 & \delta T_{oo}
            \end{array} 
            \right), \\
X^{S} &=& \left(
                  \begin{array}{c}
                     -(1+T)_{ee}^{-1} {M_{oe}}^{\dag} \hat{X}^{S}_{o} \\
                      \hat{X}^{S}_{o}
                  \end{array}
                  \right), \\
Y^{S} &=& \left(
                  \begin{array}{c}
                     -(1+T)_{ee}^{-1} M_{eo} \hat{Y}^{S}_{o} \\
                      \hat{Y}^{S}_{o}
                  \end{array}
                  \right), \\
Z^{S} &=& \left(
                  \begin{array}{c}
                     -(1+T)_{ee}^{-1} M_{eo} \hat{Y}^{S}_{o} \\
                      (1+T)_{oo}^{-1} M_{oe} (1+T)_{ee}^{-1} M_{eo} \hat{Y}^{S}_{o}
                  \end{array}
                  \right), \\
\hat{X}^{S}_{o} &=&  (1+T)_{oo}^{-1} \left(\left.\hatDS_{oo}\right.^{\dag}\right)^{-1}
                                     \left(\hatDS_{oo}\right)^{-1} \phi_{o},\\
\hat{Y}^{S}_{o} &=&  \left(\hatDS_{oo}\right)^{-1} \phi_{o}.
\end{eqnarray}
The $\gamma_{5}$ Hermiticity is slightly different for $\hatDS_{oo}$,
which is 
$\left.\hatDS_{oo}\right.^{\dag}=
\gamma_{5}(1+T)_{oo}\hatDS_{oo}(1+T)_{oo}^{-1} \gamma_{5}$.

The force contribution from the determinant of the SW term is
written as
\begin{equation}
\delta S^{S}_{\mathit{det}}=-2 \mbox{Tr}\left[\delta T (1+T)^{-1}\right],
\end{equation}
at every lattice site. 
The term {\Large $\times$} in Fig.~\ref{fig:ForceDet} is replaced by
\begin{equation}
\left(\mbox{\Large$\times$}\right)_{\mu\nu}(n)=
\left\{
-\frac{i c_{\mathrm{sw}}\kappa}{8}\mbox{tr}_{\mathrm{dirac}}
 \left[\sigma_{\mu\nu}
   Z^{S}(n)X^{S}(n)^{\dag}\right]
-\frac{i c_{\mathrm{sw}}\kappa}{8}\mbox{tr}_{\mathrm{dirac}}
 \left[\sigma_{\mu\nu} (1+T)^{-1}(n)\right]\right\}
+\mbox{H.c.}
\end{equation}

\subsection{PHMC}
In the PHMC algorithm, the term {\Large $\times$} from the SW-term
$\delta T$ in Fig.~\ref{fig:ForceDet} is written as
\begin{eqnarray}
\left(\mbox{\Large$\times$}\right)_{\mu\nu}(n)
& = &
\left\{
 \frac{i c_{\mathrm{sw}}\kappa}{8}
\sum_{j=1}^{\Npoly}
  \left(
      \mbox{tr}_{\mathrm{dirac}}
      \left[\sigma_{\mu\nu} Z^{P(j)}(n)X^{P(j)}(n)^{\dag} \right] 
  \right)
-\frac{i c_{\mathrm{sw}}\kappa}{8}\mbox{tr}_{\mathrm{dirac}}
      \left[\sigma_{\mu\nu} (1+T)^{-1}(n)\right]\right\}
\nonumber\\
&&
+\mbox{H.c.}
\end{eqnarray}

%%%%%%%%%%%%%%%%%%%%%%%%%%%%%%%%%%%%%%%%%%%%%%%%%%%%%%%%%%%%%%%%%%%%%

%%%%%%% Tables.tex

\begin{table}[H]
\caption{
\label{tab:Lattice_Parameters}
  Lattice parameters.
  }
\begin{ruledtabular}
\begin{tabular}{cccc}
             &   Small heavy   &   Large heavy   &   Large light   \\ \hline
Size         & $8^3\times 16$  & $16^3\times 48$ & $16^3\times 48$ \\
$\beta$      &  5.0            & 5.2             & 5.2     \\
$\kappa$     & 0.1415          & 0.1340          & 0.1350 \\
$c_{\rm sw}$ & 1.855           & 2.02            & 2.02   \\
$m_{\mathit{PS}}/m_{\mathit{V}}$ 
             & $\sim 0.8$\footnote{This number is measured on a $12^3\times 32$ lattice.}
             & $\sim 0.8$
             & $\sim 0.7$
\end{tabular}
\end{ruledtabular}
\end{table}

\begin{table}[H]
\caption{
\label{tab:Small_Heavy_Parameters}
  Parameters on the small heavy lattice. 
  MD step size $dt$ satisfies $dt\times\NMD=1$.
  }
\begin{ruledtabular}
\begin{tabular}{cccccc}
            &   HMC  & A-HMC  & S-HMC  & C-PHMC & A-PHMC    \\ \hline
$\NMD$
            &\begin{tabular}{c} 100, 50,\\ 40, 30         \end{tabular}
            &\begin{tabular}{c} 100, 50,\\ 40, 30,\\ 25, 20 \end{tabular}
            &\begin{tabular}{c} 100, 50,\\ 40, 32,\\ 25, 20 \end{tabular}
            &\begin{tabular}{c}  50, 40,\\ 32, 25,\\ 20     \end{tabular}
            &         32  \\ \hline
 $N_{\it poly}$ 
            & -      & -      & -      &\begin{tabular}{c}
                                         18, 20,\\ 22, 24,\\ 28, 30,\\
                                        (for $\NMD=32$)\\
                                         26,\\
                                        (for all $\NMD$).
                                        \end{tabular}
                                       &\begin{tabular}{c}
                                         14, 16,\\ 18, 20,\\ 22 \\
                                        \end{tabular}  \\ \hline
\begin{tabular}{c} Stopping condition\\ (force) \end{tabular}
            &$10^{-12}$&$10^{-12}$&$10^{-12}$& -        & -        \\
\begin{tabular}{c} Stopping condition\\ (Hamiltonian) \end{tabular}
            &$10^{-14}$&$10^{-14}$&$10^{-14}$&$10^{-14}$
  \footnote{This is used to generate a pseudofermion field 
            and global Metropolis test for the correction factor.} 
                                             &$10^{-14}$ \footnotemark[1]
\end{tabular}
\end{ruledtabular}
\end{table}

\begin{table}[H]
\caption{
\label{tab:Large_Heavy_Parameters_HMC}
  Simulation with the HMC algorithm on the large heavy lattice.
  }
\begin{ruledtabular}
\begin{tabular}{cccc}
                       & HMC           & A-HMC          & S-HMC        \\ \hline
$\NMD$                 & 160           & 100            & 80           \\
\begin{tabular}{c} Stopping condition\\ (force) \end{tabular}
                       & $10^{-18}$
\footnote{The residual is defnined by $|A x - b|$ in the HMC case.} 
                                       & $10^{-8}$      & $10^{-8}$    \\
\begin{tabular}{c} Stopping condition\\ (Hamiltonian) \end{tabular}
                       & $10^{-20}$ \footnotemark[1]
                                       & $10^{-14}$     & $10^{-14}$   \\ \hline
 Trajectories          & 3000          & 1200           & 1200         \\
$\langle dH\rangle$    & 0.144(15)     & 0.182(17)      & 0.187(28)    \\
 HMC acceptance        & 0.799(9)      & 0.764(12)      & 0.759(23)    \\
 Plaquette             & 0.52801(10)   & 0.52803(9)     & 0.52827(13) 
\end{tabular}
\end{ruledtabular}
\end{table}

\begin{table}[H]
\caption{
\label{tab:Large_Light_Parameters_HMC}
  Same as in Table~\ref{tab:Large_Heavy_Parameters_HMC} 
  but for the large light lattice.
}
\begin{ruledtabular}
\begin{tabular}{cccc}
                       & HMC           & A-HMC         & S-HMC        \\ \hline
$\NMD$                 & 200           & 125           & 100          \\
\begin{tabular}{c} Stopping condition\\ (force) \end{tabular}
                       & $10^{-18}$
\footnote{The residual is defnined by $|A x - b|$ in the HMC case.} 
                                       & $10^{-8}$     & $10^{-8}$    \\
\begin{tabular}{c} Stopping condition\\ (Hamiltonian) \end{tabular}
                       & $10^{-20}$ \footnotemark[1]
                                       & $10^{-14}$    & $10^{-14}$   \\ \hline
Trajectories           & 3000          & 1200          & 850          \\
$\langle dH\rangle$    & 0.313(23)     & 0.182(17)     & 0.218(22)    \\
 HMC acceptance        & 0.702(11)     & 0.724(13)     & 0.761(16)    \\
 Plaquette             & 0.53413(5)    & 0.53404(9)    & 0.53393(11) 
\end{tabular}
\end{ruledtabular}
\end{table}

%%%%%%%%%%%%%%%%%%%%%%%%%%%%%%
\begin{table}[H]
\caption{
\label{tab:Large_Heavy_Parameters_PHMC}
  Simulation with the C-PHMC algorithm on the large heavy
  lattice.
  }
\begin{ruledtabular}
\begin{tabular}{cccc}
                            & C-PHMC(70)  & C-PHMC(80)  & C-PHMC(90)   \\ \hline
$\NMD$                      & 80          & 80          & 80           \\
Stopping condition 
 \footnote{This is used for the generation of the pseudofermion field 
           and the calculation of the correction factor.}  
                            & $10^{-14}$  & $10^{-14}$  & $10^{-14}$   \\ \hline
 Trajectories               & 1300        & 1000        & 1000         \\
$\langle dH\rangle$         & 0.151(21)   & 0.187(22)   & 0.154(31)    \\
 HMC acceptance             & 0.775(17)   & 0.763(23)   & 0.787(21)    \\
$\langle dS\rangle$         & 0.244(32)   & 0.069(24)   & 0.013(7)     \\
Correction acceptance       & 0.731(20)   & 0.851(21)   & 0.930(13)    \\
 Total acceptance           & 0.568(17)   & 0.658(27)   & 0.732(30)    \\
 Plaquette                  & 0.52803(11) & 0.52809(10) & 0.52809(10)
\end{tabular}
\end{ruledtabular}
\end{table}

\begin{table}[H]
\caption{
\label{tab:Large_Light_Parameters_PHMC}
  Same as Table~\ref{tab:Large_Heavy_Parameters_PHMC} but for
  the large light lattice.
  } 
\begin{ruledtabular}
\begin{tabular}{cccc}
                     & C-PHMC(120) & C-PHMC(140) & C-PHMC(160)  \\ \hline
$\NMD$               & 100         & 100         & 100          \\
Stopping condition 
 \footnote{This is used for the generation of the pseudofermion field 
           and the calculation of the correction factor.}  
                     & $10^{-14}$  & $10^{-14}$  & $10^{-14}$   \\ \hline
 Trajectories        & 1600        & 1200        & 1100         \\
$\langle dH\rangle$  & 0.197(18)   & 0.243(43)   & 0.194(20)    \\
 HMC acceptance      & 0.750(12)   & 0.768(13)   & 0.765(14)    \\
$\langle dS\rangle$  & 0.719(48)   & 0.188(17)   & 0.052(14)    \\
Correction acceptance& 0.563(18)   & 0.770(12)   & 0.886(19)    \\
 Total acceptance    & 0.422(14)   & 0.597(15)   & 0.678(17)    \\
 Plaquette           & 0.53417(11) & 0.53396(18) & 0.53411(15)
\end{tabular}
\end{ruledtabular}
\end{table}

%%%%%%%%%%%%% Nf = 1+1
\begin{table}[H]
\caption{
\label{tab:CompPlaqNf1plus1}
  A comparison of the two- and $(1+1)$-flavor QCD simulations 
  at $\beta=5.0$, $8^3\times
  16$, $\kappa=0.1415$, $c_{\mathrm{sw}}=1.855$.
  } 
\begin{ruledtabular}
\begin{tabular}{cccc}
            & S-HMC          & C-PHMC(26)  & C-PHMC(26)  \\
            & $N_{f}=2$      & $N_{f}=2$   & $N_{f}=1+1$ \\ \hline
$\NMD$      & 32             & 32          & 32          \\
Trajectories
            & 5000           & 5000        & 5000        \\
$\langle dH\rangle$
            & 0.2634(107)    & 0.2236(106) & 0.1262(70)  \\
 HMC acceptance
            & 0.7172(79)     & 0.7444(70)  & 0.7994(78)  \\
$\langle dS\rangle$ (quark 1)
            &  -             & 0.0553(59)  & 0.0234(36)  \\
 Correction acceptance (quark 1)
            &  -             & 0.8595(53)  & 0.9264(54)  \\
$\langle dS\rangle$ (quark 2)
            &  -             &  -          & 0.0167(37)  \\
 Correction acceptance (quark 2)
            &  -             &  -          & 0.9370(58)  \\
 Total acceptance
            & 0.7172(79)     & 0.6398(117) & 0.6950(74)  \\ \hline 
Plaquette
            & 0.43877(22)    & 0.43839(27) & 0.43857(20)
\end{tabular}
\end{ruledtabular}
\end{table}

%%%%%%%%%%%%% Nf = 1 + 1 on Large Lattice
\begin{table}[H]
\caption{
\label{tab:CompPHMCNf1plus1_Large}
  Simulation statistics with the $N_f=1+1$ C-PHMC algorithm 
  on large lattices.
  }
\begin{ruledtabular}
\begin{tabular}{ccc}
                    & Large heavy  & Large light   \\
                    & C-PHMC(80)   & C-PHMC(140)   \\
                    & $N_{f}=1+1$  &  $N_{f}=1+1$  \\ \hline
$\NMD$              & 80           & 100           \\
 Trajectories       & 1000         & 1500          \\
$\langle dH\rangle$ & 0.081(14)    & 0.042(11)     \\
 HMC acceptance     & 0.829(14)    & 0.872(14)     \\
$\langle dS\rangle$ 
 (quark 1)          & 0.014(7)     & 0.042(10)     \\
 Correction acceptance (quark 1)
                    & 0.944(11)    & 0.878(9)      \\
$\langle dS\rangle$
 (quark 2)          & 0.0084(61)   & 0.047(10)     \\
 Correction acceptance (quark 2)
                    & 0.936(9)     & 0.876(16)     \\
 Total acceptance   & 0.733(20)    & 0.671(20)     \\ \hline
Plaquette           & 0.52782(12)  & 0.53392(9) 
\end{tabular}
\end{ruledtabular}
\end{table}

\begin{table}[H]
\caption{
\label{tab:HR_Comp_Small}
  Simulation parameters for a $(2+1)$-flavor QCD simulation.
  $\beta=4.8, 4^3\times 8, c_{\mathrm{sw}}=1.00, \kappa_{ud}=0.150,
  \kappa_{s}=0.140$ are used.  The stopping conditions are defined as follows:
  (a) the force calculation from the $N_f=2$ pseudofermion action,
  (b) the calculation of the Hamiltonian of the $N_f=2$ pseudofermion action,
  (c) the generation of the pseudofermion field, and the calculation of
     the correction factor for the single flavor part.
}
\begin{ruledtabular}
\begin{tabular}{cccc}
                  & Hybrid-R (extrapolated) & C-PHMC(10)  & C-PHMC(10)     \\
                             & $N_{f}=2+1$  & $N_{f}=2+1$ & $N_{f}=2+1$    \\ \hline
                  $\NMD$     & -            &  20         &  10            \\
Stopping condition  (a)
                             & -            & $10^{-14}$  & $10^{-14}$     \\
Stopping condition  (b)
                             & -            & $10^{-14}$  & $10^{-14}$     \\
Stopping condition  (c)
                             & -            & $10^{-14}$  & $10^{-14}$     \\ \hline
$\langle dH\rangle$          & -            & 0.055(5)    & 0.839(28)      \\
HMC acceptance ratio         & -            & 0.877(7)    & 0.521(12)      \\
$\langle dS\rangle$          & -            & 0.00014(61) & 0.00056(62)    \\
Correction acceptance ratio  & -            & 0.9843(21)  & 0.9861(22)     \\
Total acceptance ratio       & -            & 0.864(7)    & 0.514(12)      \\ \hline
Plaquette                    & 0.39702(13)  & 0.39669(38) & 0.39695(32)
\end{tabular}
\end{ruledtabular}
\end{table}

\begin{table}[H]
\caption{
\label{tab:HR_Comp_Middle}
  Simulation parameters for a $(2+1)$-flavor QCD simulation.
  $\beta=5.0, 8^3\times 16, c_{\mathrm{sw}}=2.08, \kappa_{ud}=0.1338,
   \kappa_{s}=0.1330$ are used. The definition of the stopping condition
  (a)-(c) is the same as those in Table~\ref{tab:HR_Comp_Small}. 
}
\begin{ruledtabular}
\begin{tabular}{ccc}
                             & Hybrid-R (extrapolated) & C-PHMC(58)     \\
                             & $N_{f}=2+1$  & $N_{f}=2+1$    \\ \hline
                  $\NMD$     & -            &  32            \\
Stopping condition (a)
                             & -            & $10^{-9}$      \\
Stopping condition (b)
                             & -            & $10^{-14}$     \\
Stopping condition (c)
                             & -            & $10^{-14}$     \\ \hline
$\langle dH\rangle$          & -            & 0.194(9)       \\
HMC acceptance ratio         & -            & 0.743(7)       \\
$\langle dS\rangle$          & -            & 0.019(3)       \\
Correction acceptance ratio  & -            & 0.926(5)       \\
Total acceptance ratio       & -            & 0.688(8)       \\ \hline
Plaquette                    & 0.53161(7)   & 0.53145(11)    
\end{tabular}
\end{ruledtabular}
\end{table}

%%%%%%%%%%% Figures.tex
\clearpage

\newcommand{\figscale}{0.9}

%%%%%%%%%%%%%%% Compare UPU and PUP in S-HMC
\begin{figure}[H]
\centering
\includegraphics[scale=\figscale]{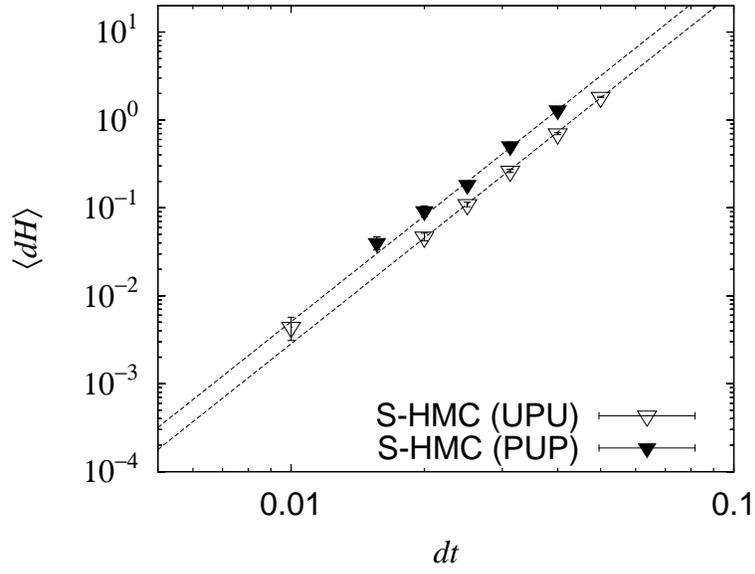}
\caption{
\label{fig:dHvsdt_UPUvsPUP}
  MD step size dependence of $\langle dH \rangle$ for two integration
  methods ($UPU$ and $PUP$) in the MD evolution.
  The lines show the fit with $\langle dH\rangle = \pi (a\cdot dt)^4$.
  }
\end{figure}

\begin{figure}[H]
\centering
\includegraphics[scale=\figscale]{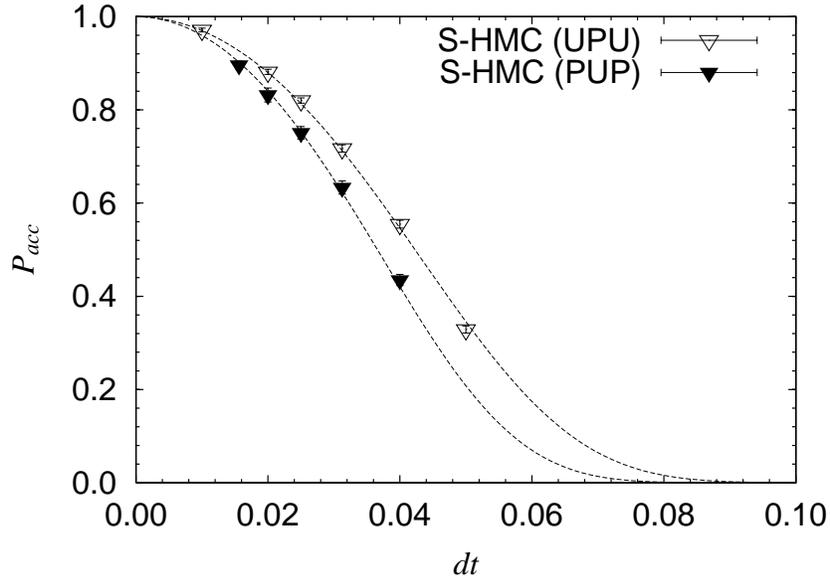}
\caption{
\label{fig:Paccvsdt_UPUvsPUP}
  MD step size dependence of the acceptance for two integration
  methods ($UPU$ and $PUP$) in the MD evolution.
  The lines show the function 
  ${\mathrm{erfc}}[\sqrt{\pi}(a\cdot dt)^2/2]$ 
  with $a$ obtained from Fig.~\ref{fig:dHvsdt_UPUvsPUP}.
  }
\end{figure}

%%%%%%%%%%%%%%% Compare HMC, A-HMC, S-HMC
\begin{figure}[H]
\centering
\includegraphics[scale=\figscale]{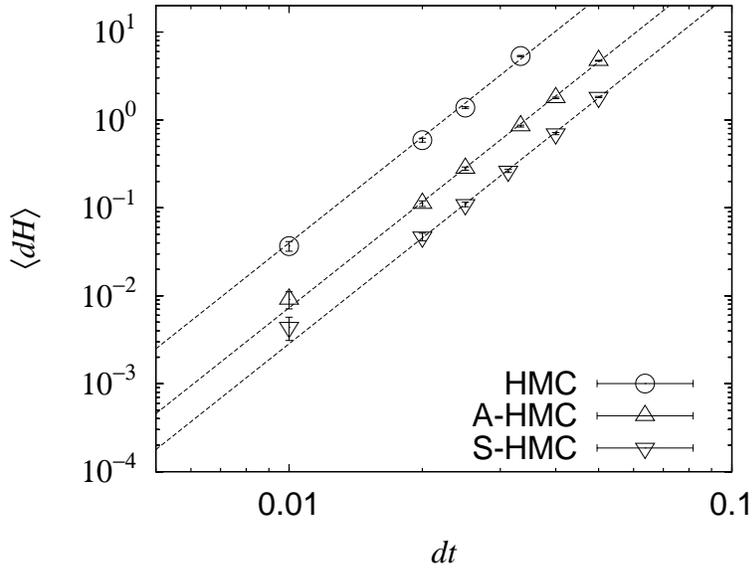}
\caption{
\label{fig:dHvsdt}
  MD step size dependence of $\langle dH \rangle$ for preconditioned
  and unpreconditioned effective actions.
  The lines show the fit with $\langle dH\rangle = \pi (a\cdot dt)^4$.}
\end{figure}

\begin{figure}[H]
\centering
\includegraphics[scale=\figscale]{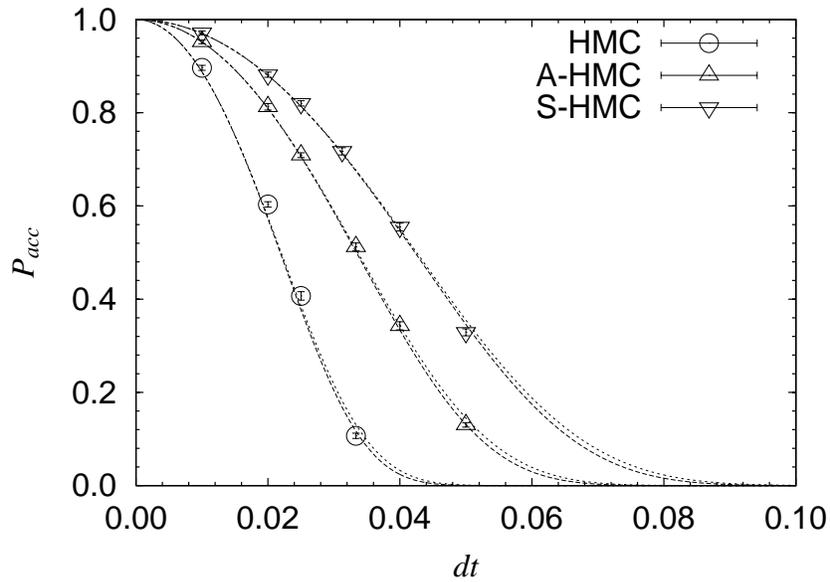}
\caption{
\label{fig:Paccvsdt}
  MD step size dependence of the acceptance for preconditioned
  and unpreconditioned effective actions.
  The dashed lines show the function 
  ${\mathrm{erfc}}[\sqrt{\pi}(a\cdot dt)^2/2]$ 
  with $a$ obtained from Fig.~\ref{fig:dHvsdt}.
  The dotted lines are approximations
  $\exp\left[-(a\cdot dt)^2 - (a\cdot dt)^4/2 \right]$.
  }
\end{figure}

\begin{figure}[H]
\centering
\includegraphics[scale=\figscale]{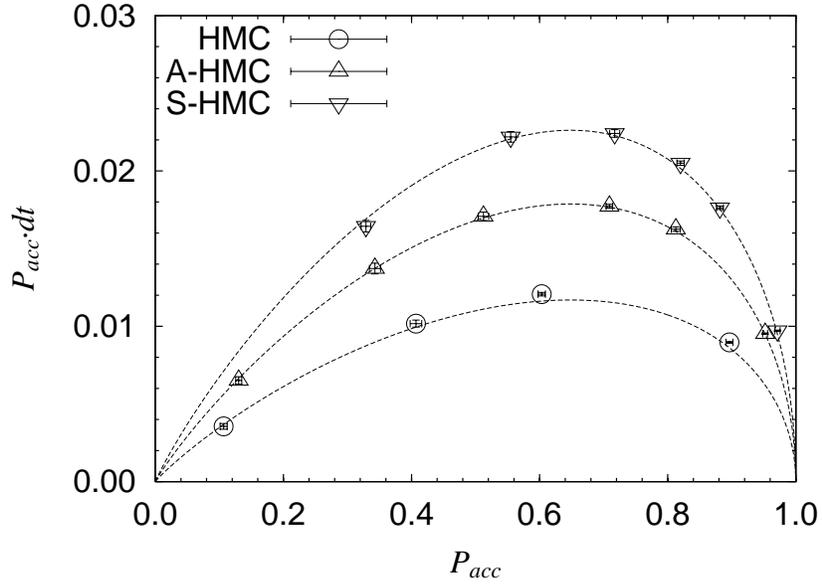}
\caption{
\label{fig:EffvsPacc}
  Efficiency $P_{\mathit{acc}}\cdot dt$.
  The lines show the function
  $P_{\mathit{acc}}\sqrt{\sqrt{1-2\log P_{\mathit{acc}}}-1}/a$
  with $a$ obtained in Fig.~\ref{fig:dHvsdt}.
  }
\end{figure}

%%%%%%%%%%%%%%% Large Lattice Reversibility for HMC
\begin{figure}[H]
\centering
\includegraphics[scale=\figscale]{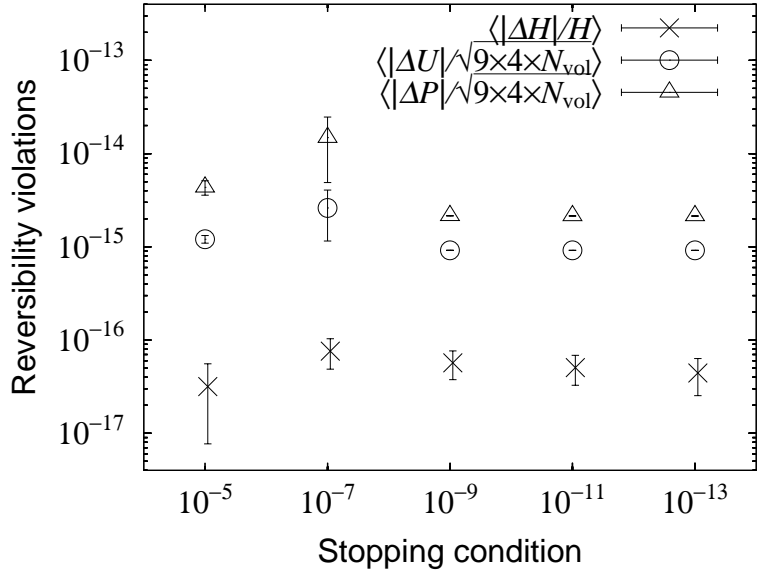}
\caption{
\label{fig:DHDUDPrev_LH_HMC}
  The violation of the reversibility as a function of the stopping
  condition on the large heavy lattice.
  }
\end{figure}

\begin{figure}[H]
\centering
\includegraphics[scale=\figscale]{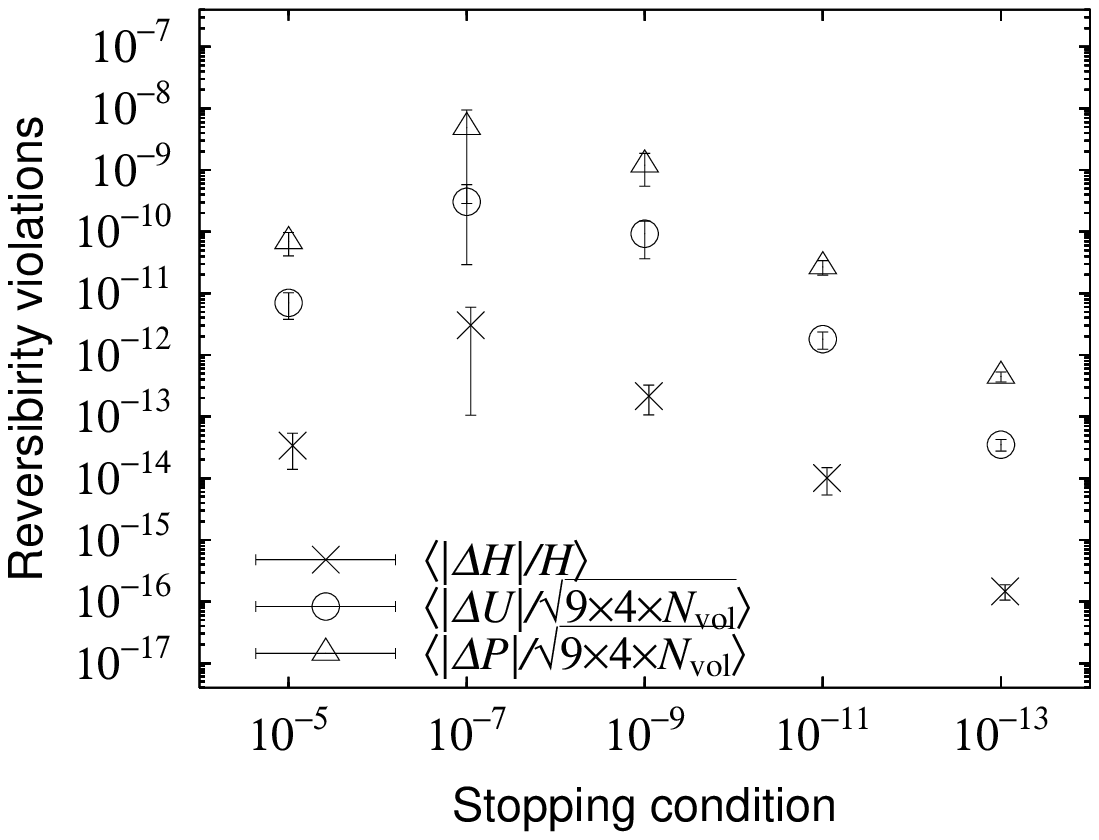}
\caption{
\label{fig:DHDUDPrev_LL_HMC}
  Same as Fig.~\ref{fig:DHDUDPrev_LH_HMC} but for the 
  large light lattice.
  }
\end{figure}

%%%%%%%%%%%%%%% PHMC on small lattice

\begin{figure}[H]
\centering
\includegraphics[scale=\figscale]{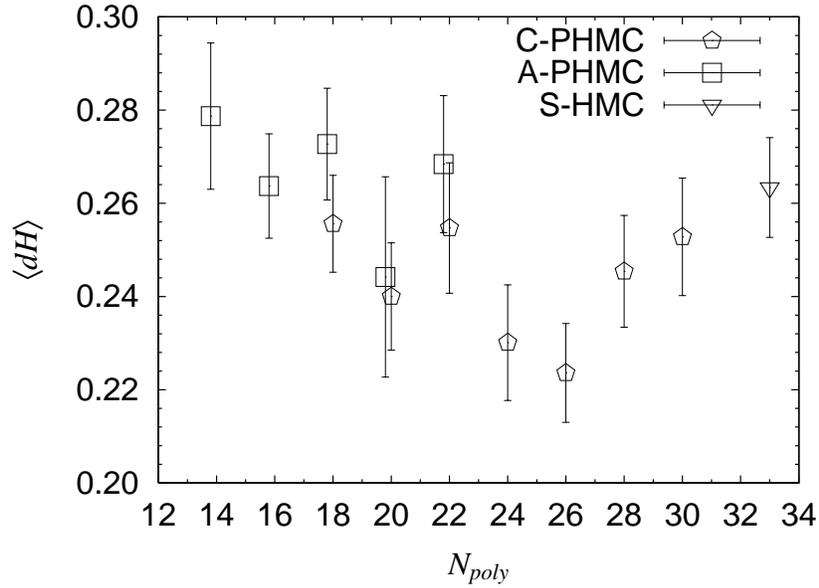}
\caption{
\label{fig:dHvsNpoly_PHMC}
  $\Npoly$ dependence of $\langle dH\rangle$
  on the small heavy lattice.
  $\langle dH\rangle$ with the S-HMC algorithm is also
  plotted on the most right side in the figure
  for comparison.
  }
\end{figure}

\begin{figure}[H]
\centering
\includegraphics[scale=\figscale]{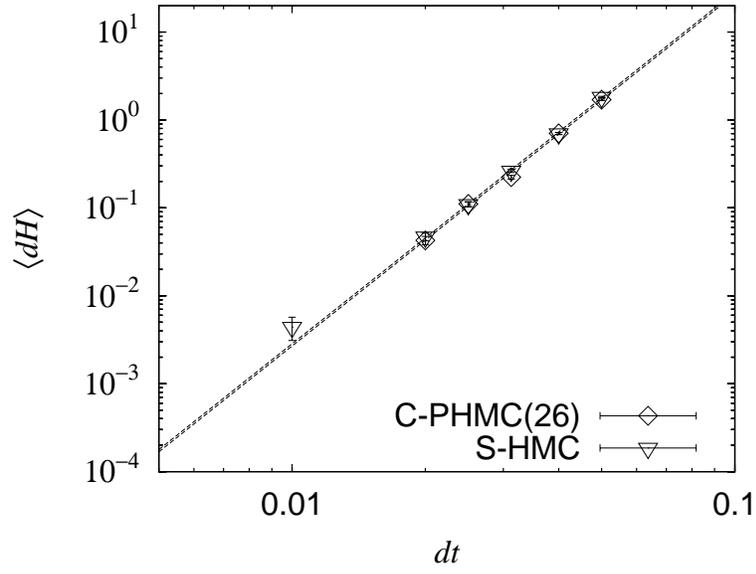}
\caption{
\label{fig:dHvsdt_PHMC}
  $\langle dH\rangle$ versus $dt$ with C-PHMC(26) and S-HMC
  algorithms.
  }
\end{figure}

\begin{figure}[H]
\centering
\includegraphics[scale=\figscale]{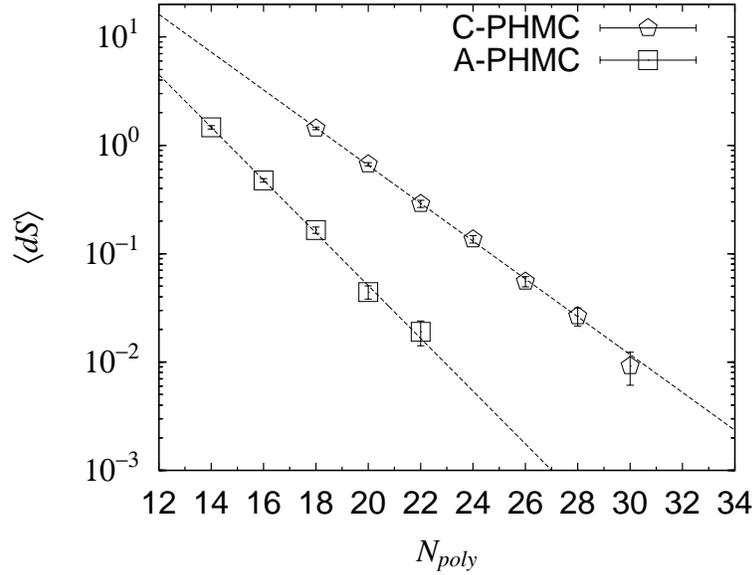}
\caption{
\label{fig:dSvsNpoly_PHMC}
  $\langle dS\rangle$ versus $\Npoly$ in the PHMC algorithm. 
  The lines show a fit function 
  $\langle dS\rangle=\pi a^{2}\exp(-2 b \Npoly)$.
  }
\end{figure}

\begin{figure}[H]
\centering
\includegraphics[scale=\figscale]{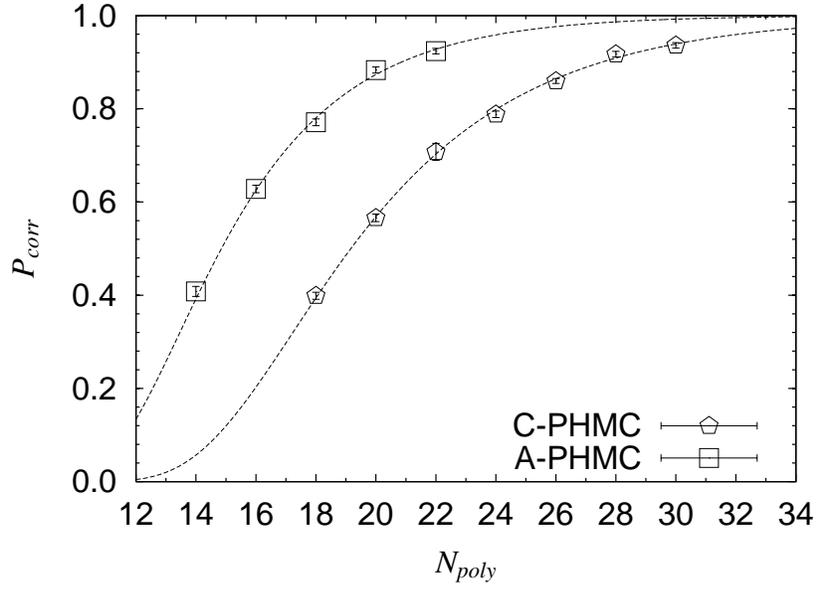}
\caption{
\label{fig:PGMPaccvsNpoly_PHMC}
  $P_{\mathit{corr}}$ as a function of $\Npoly$.
  The lines represent
  $P_{\mathit{corr}}={\mathrm{erfc}}[\sqrt{\pi}a\exp(-b\Npoly)/2]$ 
  with $a$ and $b$ are obtained from a fit in
  Fig.~\ref{fig:dSvsNpoly_PHMC}.
  }
\end{figure}

\begin{figure}[H]
\centering
\includegraphics[scale=\figscale]{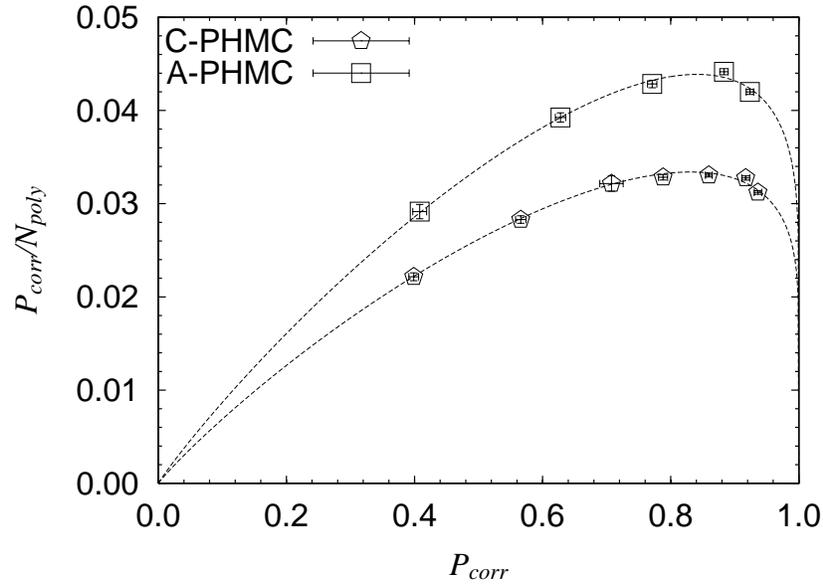}
\caption{
\label{fig:EffvsPGMPacc_PHMC}
  Efficiency of the noisy Metropolis test $P_{\mathit{corr}}/\Npoly$.
  See the text for details.
  }
\end{figure}

%%%%%%%%%%%%%%% PHMC on Large Lattice
\begin{figure}[H]
\centering
\includegraphics[scale=\figscale]{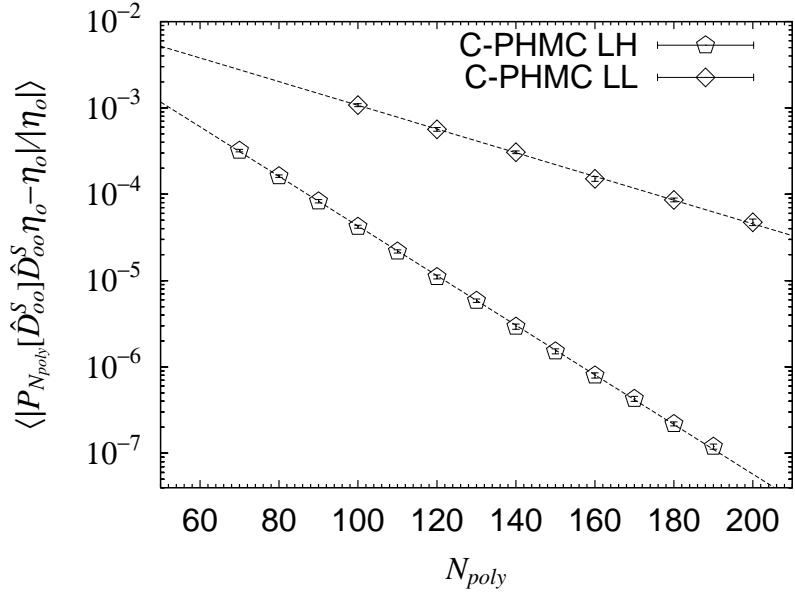}
\caption{
\label{fig:Conv_PHMC}
  $\Npoly$ dependence of the residual 
  $\langle | P_{\Npoly}[\hatDS_{oo}]\hatDS_{oo}\eta_{o}-\eta_{o}|/|\eta_{o}|\rangle$.
  }
\end{figure}

%%%%% rev check
\begin{figure}[H]
\centering
\includegraphics[scale=\figscale]{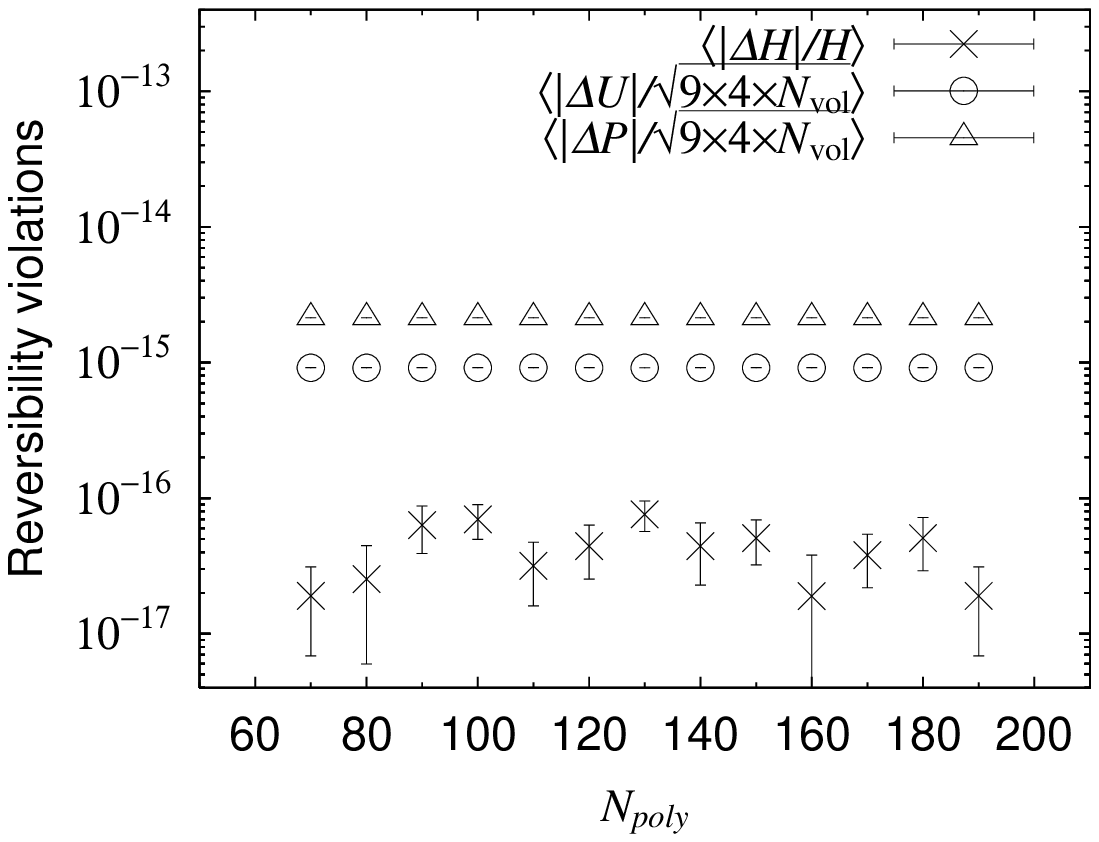}
\caption{
\label{fig:DHDUDPrev_LH_PHMC}
  $\Npoly$ dependence of the reversibility violation on the large
  heavy lattice.
  }
\end{figure}

\begin{figure}[H]
\centering
\includegraphics[scale=\figscale]{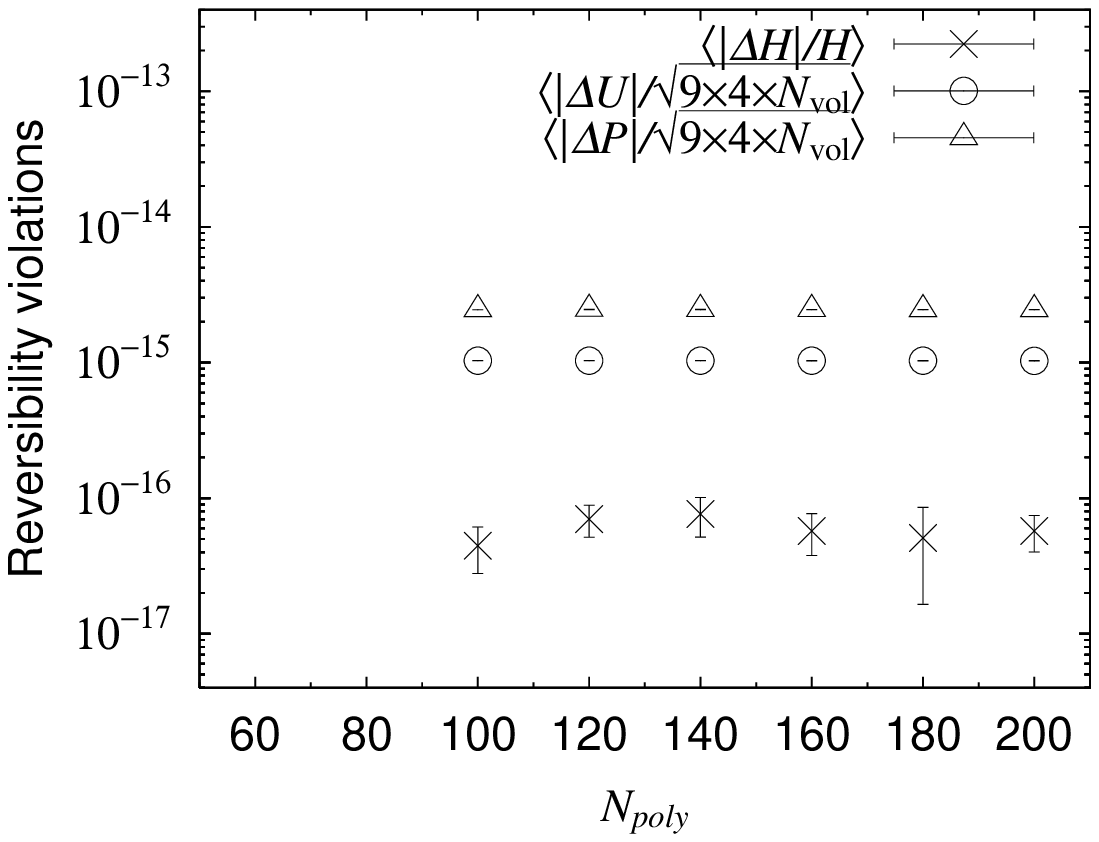}
\caption{
\label{fig:DHDUDPrev_LL_PHMC}
  Same as Fig.~\ref{fig:DHDUDPrev_LH_PHMC} but for the large light
  lattice. 
  }
\end{figure}

%%%%% Simulation result on global Metropolis test
\begin{figure}[H]
\centering
\includegraphics[scale=\figscale]{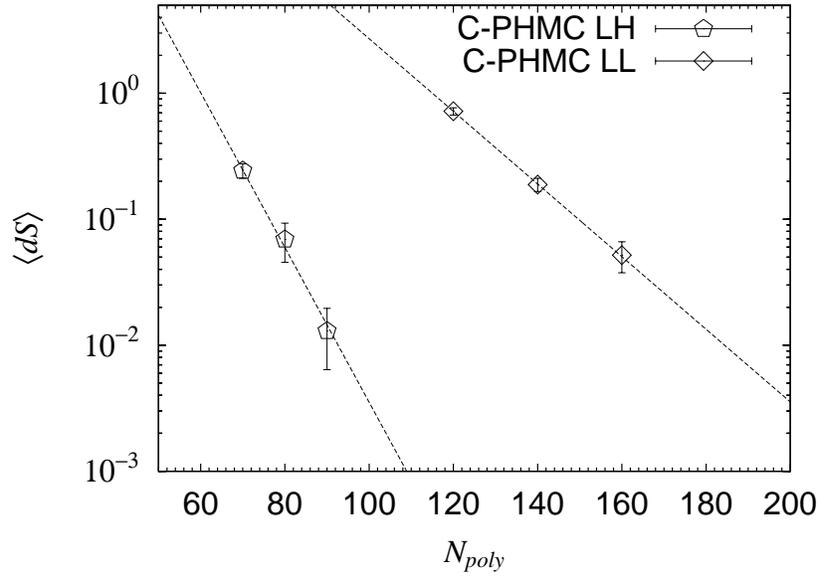}
\caption{
\label{fig:dSvsNpoly_LH_PHMC}
  $\langle dS\rangle$ versus $\Npoly$ for the large haevy
  (pentagons) and large light (diamonds) lattices.
  }
\end{figure}

\begin{figure}[H]
\centering
\includegraphics[scale=\figscale]{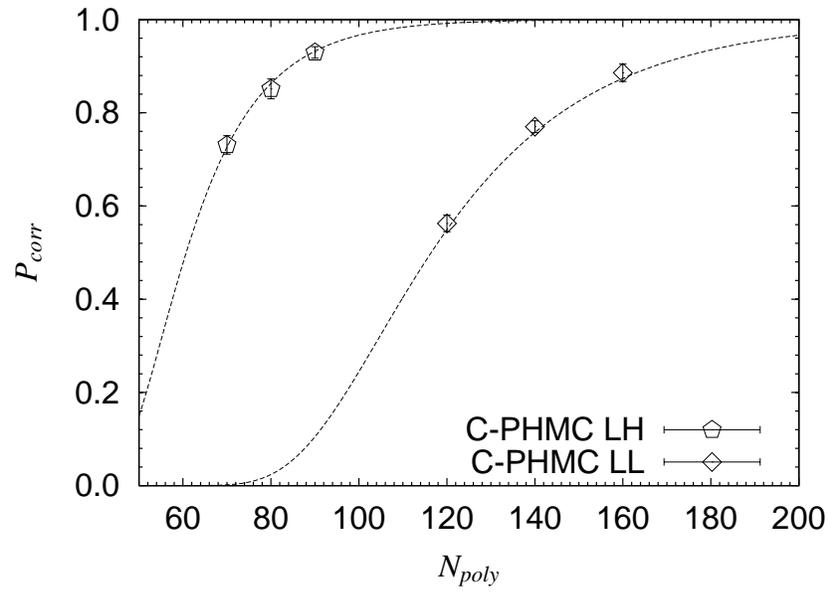}
\caption{
\label{fig:PGMPaccvsNpoly_LH_PHMC}
  $P_{\mathit{corr}}$ versus $\Npoly$ for the large haevy
  (pentagons) and large light (diamonds) lattices.
  }
\end{figure}

%%%%% efficiency
\begin{figure}[H]
\centering
\includegraphics[scale=\figscale]{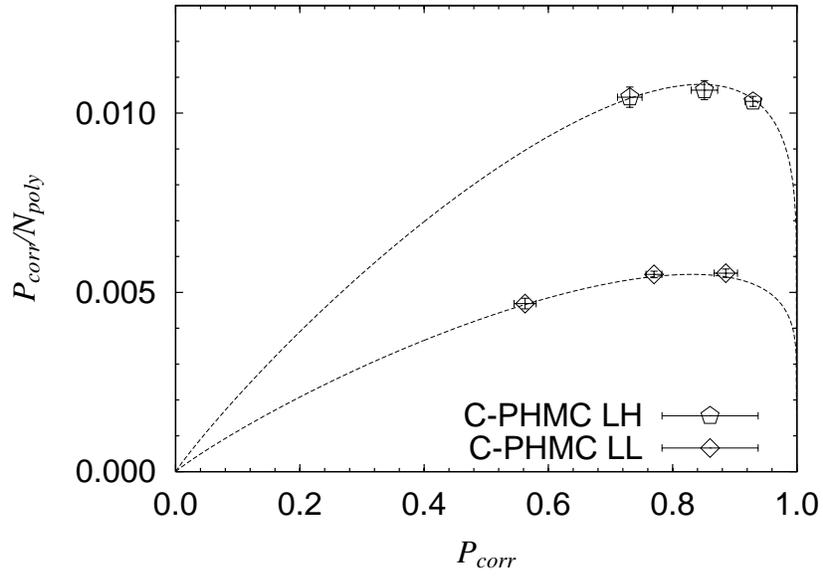}
\caption{
\label{fig:EffvsPGMPacc_LH_PHMC}
  Efficiency $P_{\mathit{corr}}/\Npoly$ versus $P_{\mathit{corr}}$ 
  on the large heavy (pentagons) and large light (diamonds)
  lattices. 
  }
\end{figure}

\begin{figure}[H]
\centering
\includegraphics[scale=\figscale]{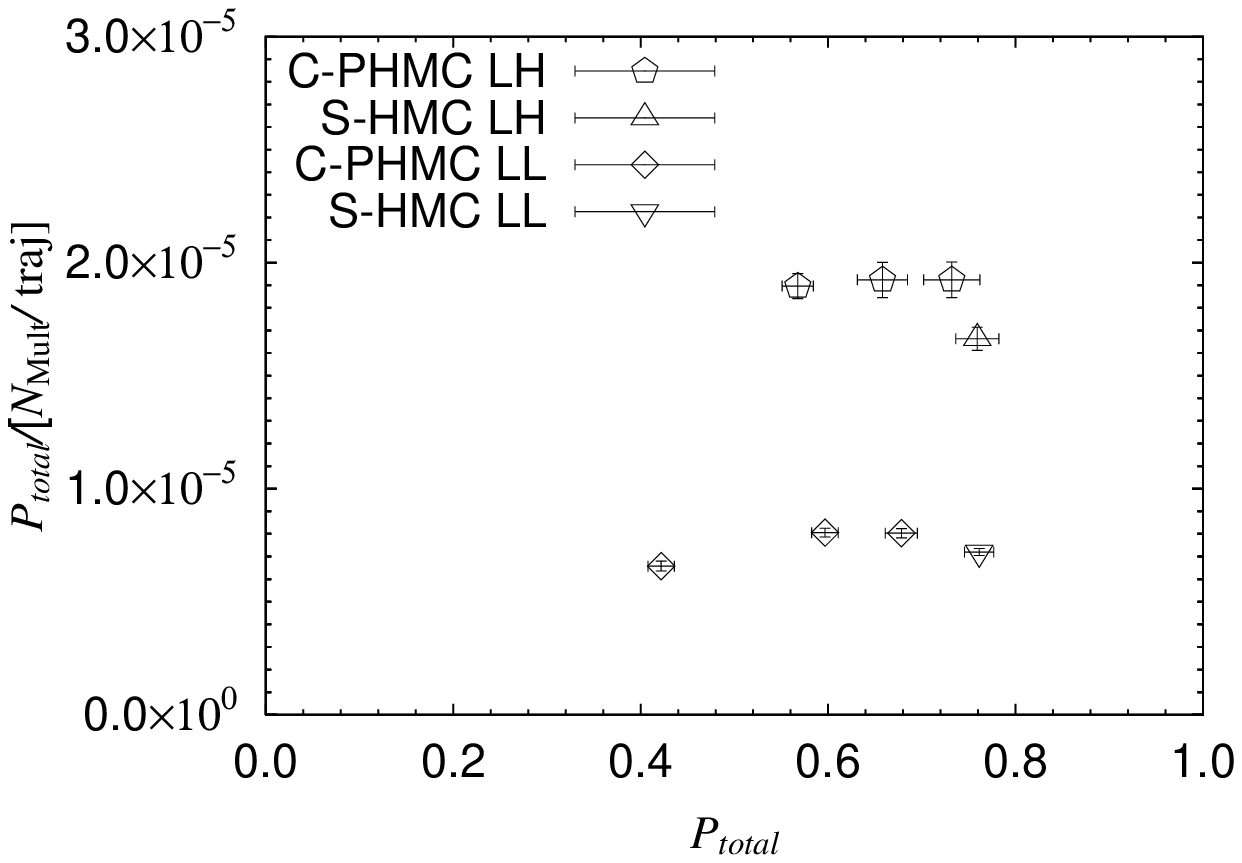}
\caption{
\label{fig:PtotaccovMult_LH_PHMC}
  Total efficiency $P_{\mathit{total}}/[\NMult/\mathrm{traj}]$ as a function
  of $P_{\mathit{total}}$.
  }
\end{figure}

%%%%%%%%%%%% N_f= 1 + 1 reversibility and convergence of polynomial, correction factor.
\begin{figure}[H]
\centering
\includegraphics[scale=\figscale]{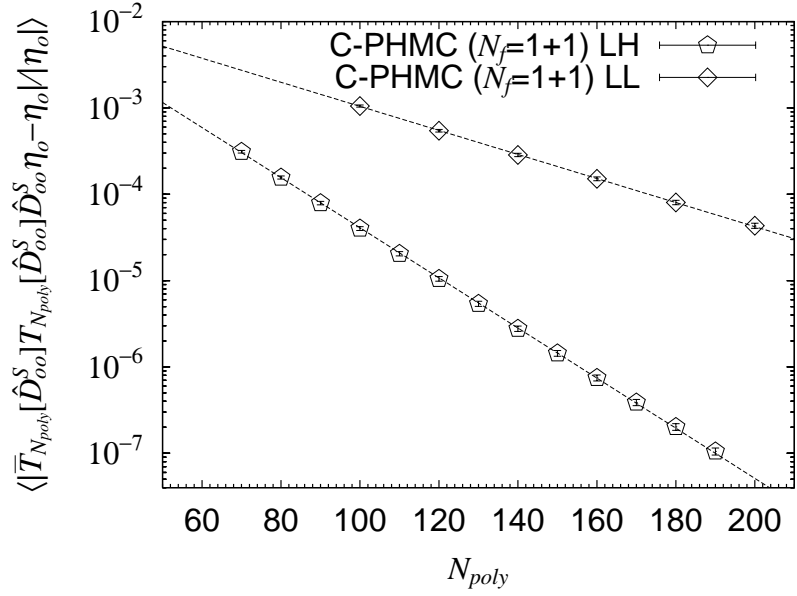}
\caption{
\label{fig:Conv_PHMC_Nf1plus1}
  $\Npoly$ dependence of the residual 
  $\langle |\overline{T}_{\Npoly}[\hatDS_{oo}]T_{\Npoly}[\hatDS_{oo}]\eta_{o}
  -\eta_{o}|/|\eta_{o}|\rangle$ ($N_f=1+1$).
  }
\end{figure}

\begin{figure}[H]
\centering
\includegraphics[scale=\figscale]{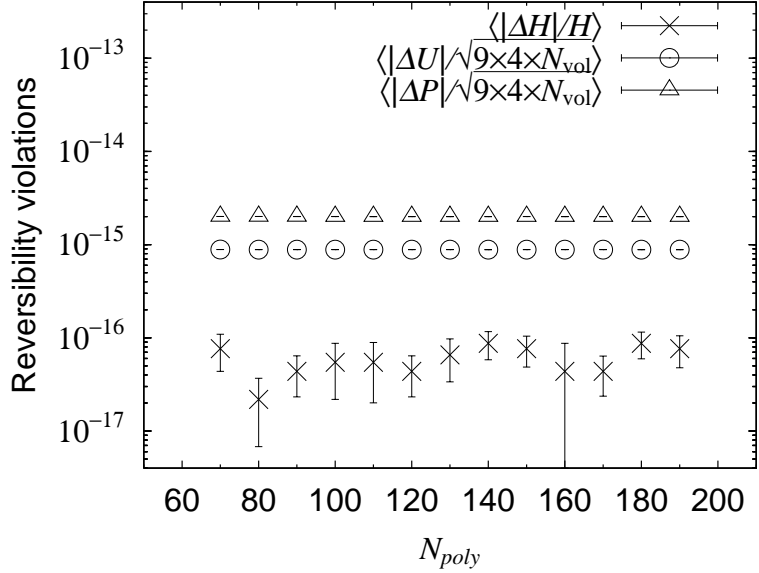}
\caption{
\label{fig:DHDUDPrev_LH_PHMC_Nf1plus1}
  $\Npoly$ dependence of the reversibility violation on the large
  heavy lattice ($N_f=1+1$).
}
\end{figure}

\begin{figure}[H]
\centering
\includegraphics[scale=\figscale]{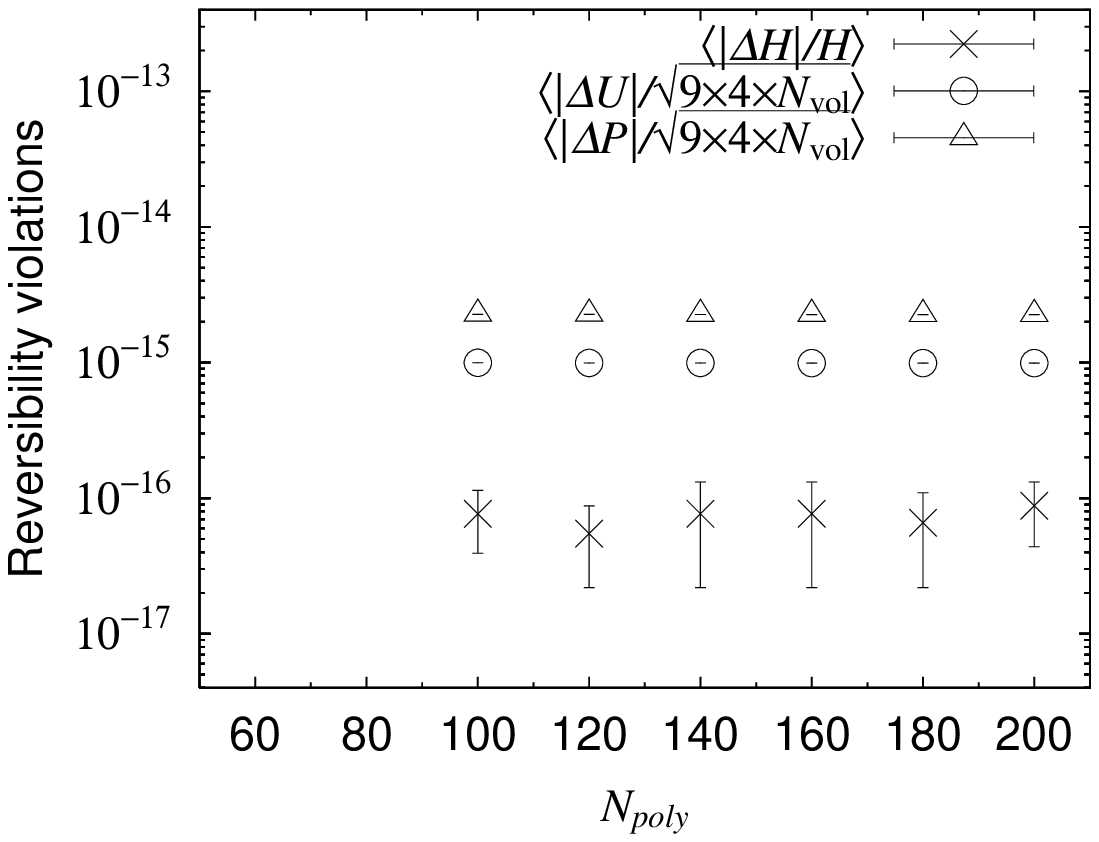}
\caption{  
\label{fig:DHDUDPrev_LL_PHMC_Nf1plus1}
  Same as Fig.~\ref{fig:DHDUDPrev_LH_PHMC_Nf1plus1} but for the large light
  lattice. 
}
\end{figure}

\begin{figure}[H]
\centering
\includegraphics[scale=\figscale]{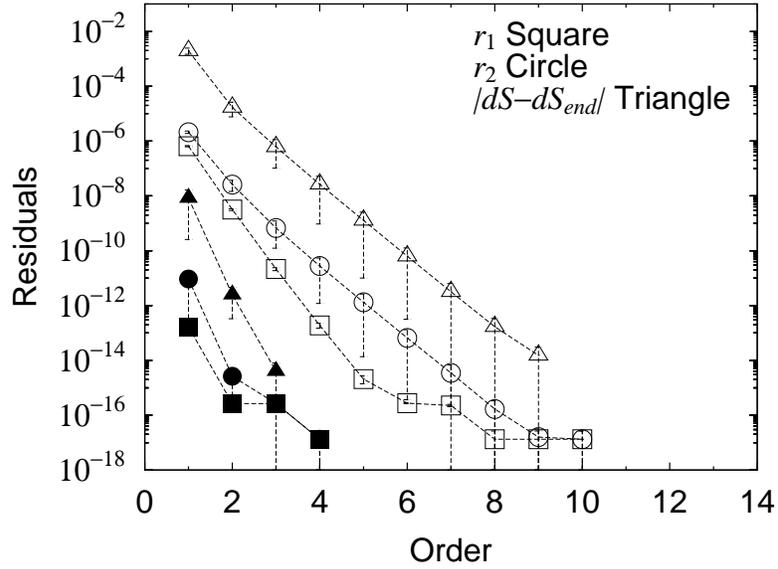}
\caption{
\label{fig:dSconv_LH_PHMC_Nf1plus1}
  Convergence behavior of the Taylor expansion of the correction matrix on 
  the large heavy lattice. Open: $\Npoly=70$; filled: $\Npoly=190$.
  }
\end{figure}

\begin{figure}[H]
\centering
\includegraphics[scale=\figscale]{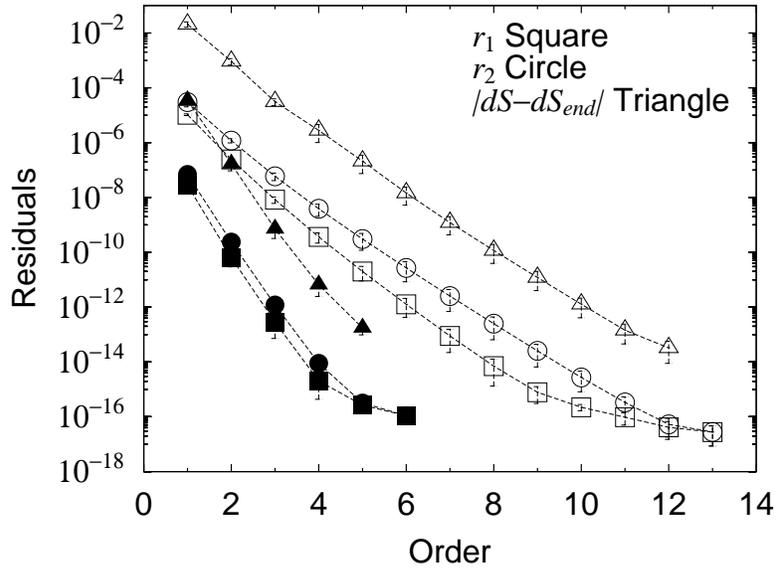}
\caption{
\label{fig:dSconv_LL_PHMC_Nf1plus1}
  Same as Fig.~\ref{fig:dSconv_LH_PHMC_Nf1plus1} but for the
  large light lattice. Open: $\Npoly=100$; filled: $\Npoly=200$.
  }
\end{figure}

%%%%%%%%%%%% Hybrid-R comparison
\begin{figure}[H]
\centering
\includegraphics[scale=\figscale]{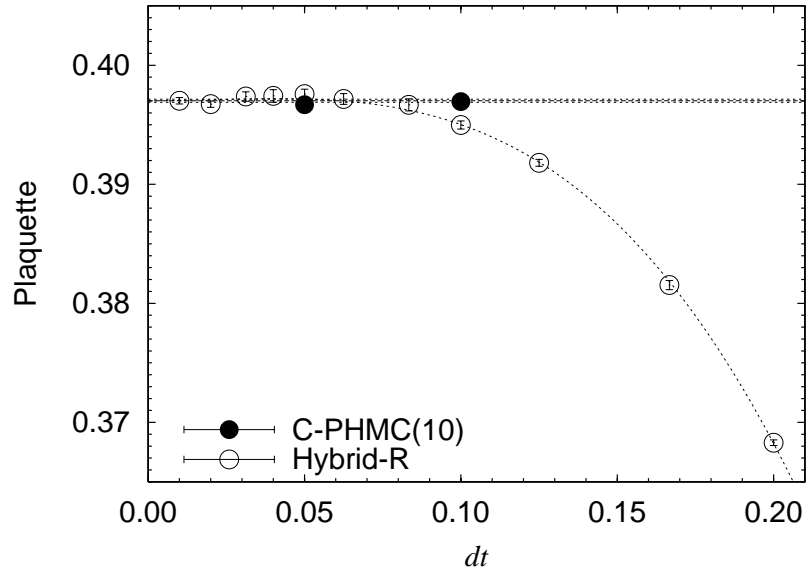}
\caption{
\label{fig:HR_Comp_Small}
  MD step size $dt$ dependence of the plaquette expectation value
  on the lattice of size $4^3\times 8$ at $\beta=4.8$, 
  $c_{\mathrm{SW}}=1.00$, $\kappa_{ud}=0.150$, $\kappa_{s}=0.140$.
  Open circles are results of the $R$ algorithm, and the filled circles
  are from our exact algorithm. }
\end{figure}

\begin{figure}[H]
\centering
\includegraphics[scale=\figscale]{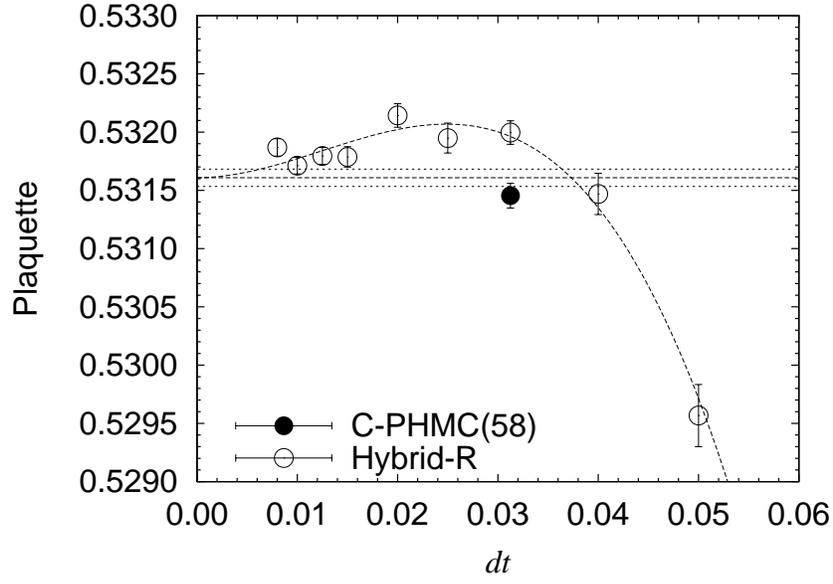}
\caption{
\label{fig:HR_Comp_Middle}
  MD step size $dt$ dependence of the plaquette expectation value
  on the lattice of size $8^3\times 16$ at $\beta=5.0$,
  $c_{\mathrm{SW}}=2.08$, $\kappa_{ud}=0.1338$, $\kappa_{s}=0.1330$.
  Open circles are results of the $R$ algorithm, and the filled circle
  is from our exact algorithm.
  }
\end{figure}

%%%%%%%%%%%% Appendix
\renewcommand{\figscale}{0.6}

\begin{figure}[H]
\centering
\includegraphics[scale=\figscale]{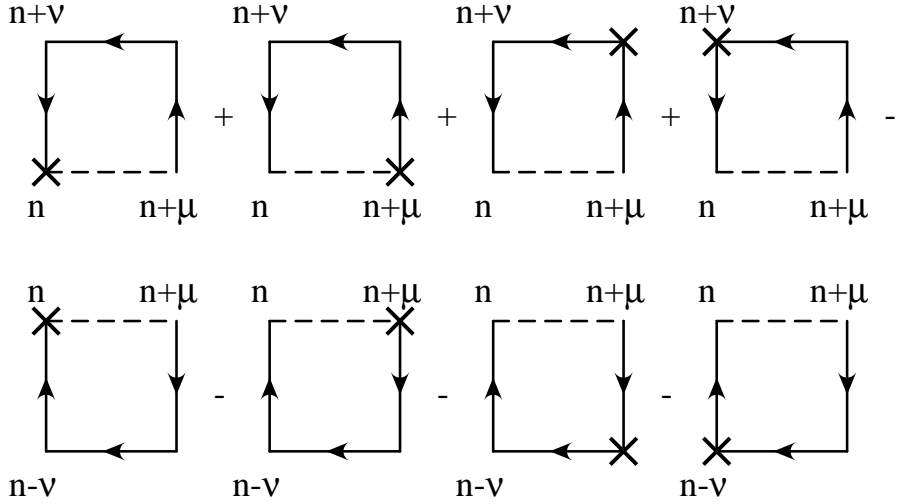}
\caption{
\label{fig:ForceDet}
  Diagrams contributing to $F_{\mu}(n)$ from the SW term.}
\end{figure}

\end{document}